\documentclass[twocolumn, 10pt]{IEEEtran}
\usepackage{graphicx}
\usepackage{amssymb}
\usepackage{amsmath}
\usepackage{mathtools}
\usepackage{dsfont}
\usepackage{cite}
\usepackage{stfloats}
\usepackage{subfigure}
\usepackage{psfrag}
\usepackage[mathscr]{euscript}
\usepackage{algorithm, algorithmic}
\usepackage{acronym}  % make an acronym
\usepackage{booktabs}
\usepackage{bbm}

\usepackage{multirow}

\usepackage{float}
\usepackage{balance}

\acrodef{CCDF}{complementary cumulative distribution function}
\acrodef{CF}{characteristic function}
\acrodef{PPP}{Poisson point processe}
\acrodef{RV}{random variable}
\acrodef{i.i.d.}{independent and identically distributed}
\acrodef{PDF}{probability distribution function}
\acrodef{CDF}{cumulative distribution function}
\acrodef{ch.f.}{characteristic function}
\acrodef{AWGN}{additive white Gaussian noise}
\acrodef{SNR}{signal-to-noise ratio}
\acrodef{LRT}{likelihood ratio test}
\acrodef{DRT}{distance ratio test}
\acrodef{GLRT}{generalized likelihood ratio test}
\acrodef{CRLB}{Cram\'{e}r-Rao lower bound}
\acrodef{CRB}{Cram\'{e}r-Rao bound}
\acrodef{ZZLB}{Ziv-Zakai lower bound}
\acrodef{ZZB}{Ziv-Zakai bound}
\acrodef{LOS}{line-of-sight}
\acrodef{ToF}{time-of-flight}
\acrodef{NLOS}{non-line-of-sight}
\acrodef{GDOP}{geometric dilution of precision}
\acrodef{GPS}{Global Positioning System}
\acrodef{FIM}{Fisher information matrix}
\acrodef{PEB}{position error bound}
\acrodef{SPEB}{squared position error bound}
\acrodef{TOA}{time-of-arrival}
\acrodef{TOF}{time-of-flight}
\acrodef{WSN}{wireless sensor network}
\acrodef{MAC}{medium access control}
\acrodef{RSS}{received signal strength}
\acrodef{WAF}{wall attenuation factor}
\acrodef{TDOA}{time difference-of-arrival}
\acrodef{RF}{radiofrequency}
\acrodef{RTT}{round-trip time}
\acrodef{AOA}{angle-of-arrival}
\acrodef{MF}{matched filter}
\acrodef{ED}{energy detector}
\acrodef{ML}{maximum likelihood}
\acrodef{MSE}{mean-square error}
\acrodef{RMSE}{root-mean-square error}
\acrodef{LEO}{localization error outage}
\acrodef{ppm}{part-per-million}
\acrodef{ACK}{acknowledge}
\acrodef{UWB}{Ultrawide bandwidth}
\acrodef{TNR}{threshold-to-noise ratio}
\acrodef{LS}{least squares}
\acrodef{IR-UWB}{impulse radio UWB}
\acrodef{FCC}{Federal Communications Commission}
\acrodef{TH}{time-hopping}
\acrodef{PPM}{pulse position modulation}
\acrodef{MUI}{multi-user interference}
\acrodef{PDP}{power delay profile}
\acrodef{BPZF}{band-pass zonal filter}
\acrodef{SIR}{signal-to-interference ratio}
\acrodef{SINR}{signal-to-interference-plus-noise ratio}
\acrodef{RFID}{radio frequency identification}
\acrodef{WPAN}{wireless personal area network}
\acrodef{WWB}{Weiss-Weinstein bound}
\acrodef{DP}{direct path}
\acrodef{MF}{matched filter}
\acrodef{MMSE}{minimum-mean-square-error}
\acrodef{SBS}{serial backward search}
\acrodef{SBSMC}{serial backward search for multiple clusters}
\acrodef{NBI}{narrowband interference}
\acrodef{WBI}{wideband interference}
\acrodef{INR}{interference-to-noise ratio}
\acrodef{CR}{channel response}
\acrodef{CIR}{channel impulse response}
\acrodef{CR}{channel  response}
\acrodef{RADAR}{radar}
\acrodef{MUR}{Multistatic radar}
\acrodef{JBSF}{jump back and search forward}
\acrodef{HDSA}{high-definition situation-aware}
\acrodef{RRC}{root raised cosine}
\acrodef{ST}{simple thresholding}
\acrodef{BTB}{Bellini-Tartara bound}
\acrodef{P-Max}{$P$-Max}  %suggestion, use with \acl{P-Max}
\acrodef{MIMO}{multiple-input multiple-output}
\acrodef{MAP}{maximum a posteriori}
\acrodef{FG}{factor graph}
\acrodef{OP}{outage probability}
\acrodef{WED}{wall extra delay}
\acrodef{RMS}{root mean square}
\acrodef{SPAWN}{sum-product algorithm over a wireless network}
\acrodef{MDD}{minimum distance distribution}
\acrodef{MAP}{maximum a posteriori probability}
\acrodef{SAP}{small cell access point}
\acrodef{UE}{user equipment}
\acrodef{MBS}{macro cell base station}
\acrodef{UER}{\ac{UE} Relay}
\acrodef{D2D}{device-to-device}
\acrodef{MBS}{macro base station}
\acrodef{CSI}{channel state information}
\acrodef{OGR}{outage guard region}
\acrodef{FUR}{feasible UER region}
\acrodef{EHR}{energy harvesting region}
\acrodef{EH}{energy harvesting}
\acrodef{D2D-EHSN}{D2D communication provided \ac{EH} small cell network}
\acrodef{D2D-EHHN}{D2D communication provided \ac{EH} heterogeneous network}
\acrodef{3GPP}{3rd Generation Partnership Project}
\acrodef{BS}{base station}
\acrodef{DF}{decode and forward}
\acrodef{CCDF}{complementary cumulative distribution function}
\acrodef{ZF}{zero forcing}
\acrodef{RZF}{regularized zero forcing}
\acrodef{WLLN}{weak law of large number}
\acrodef{SLLN}{strong law of large numbers}
\acrodef{TDD}{Time-division duplex}
\acrodef{EE}{energy efficiency} 
\acrodef{HetNet}{heterogeneous network} 
\acrodef{SCP}{Single Cell Processing}
\acrodef{CBF}{Coordinated Beamforming}
\usepackage{color}
\usepackage{dsfont}
\usepackage{bbm}

\DeclareMathAlphabet{\mathsf}{OML}{cmbr}{m}{it}

\newtheorem{definition}{\bf Definition}
\newtheorem{theorem}{\bf Theorem}
\newtheorem{lemma}{\bf Lemma}
\newtheorem{corollary}{\bf Corollary}

\newtheorem{remark}{\bf Remark}

\newtheorem{assumption}{\bf Assumption}

\newcommand{\bd}{\begin{description}}
\newcommand{\ed}{\end{description}}
\newcommand{\be}{\begin{enumerate}}
\newcommand{\ee}{\end{enumerate}}
\newcommand{\bi}{\begin{itemize}}
\newcommand{\ei}{\end{itemize}}
\newcommand{\bl}{\begin{list}}
\newcommand{\el}{\end{list}}
\newcommand{\bt}{\begin{tabbing}}
\newcommand{\et}{\end{tabbing}}

\newcommand{\paperTitle}{ Server Free Wireless Federated Learning: Architecture, Algorithm, and Analysis }

\begin{document}

{
\title{\paperTitle}

\author{
    
    Howard H. Yang, \textit{Member, IEEE},
    Zihan~Chen, \textit{Student Member, IEEE},
    % Chris~Zheng,
    and Tony Q. S. Quek, \textit{Fellow, IEEE}

% \thanks{Manuscript received Mar. 23, 2017, revised Jul. 10, 2017 and accepted Aug. 6, 2017. The associate editor coordinating the review of this letter and
%     approving it for publication was Dr. Koji Yamamoto.

%     This work was supported in part by the Zhejiang Provincial Public Technology Research of China under Grant 2016C31063 and the MOE ARF Tier 2 under Grant MOE2015-T2-2-104.}

\thanks{H. H. Yang is with the Zhejiang University/University of Illinois at Urbana-Champaign Institute, Zhejiang University, Haining 314400, China, the College of Information Science and Electronic Engineering, Zhejiang University, Hangzhou 310007, China, and the Department of Electrical and Computer Engineering, University of Illinois at Urbana-Champaign, Champaign, IL 61820, USA (email: haoyang@intl.zju.edu.cn).}

\thanks{Z. Chen and T.~Q.~S.~Quek are with the Information Systems Technology and Design Pillar, Singapore University of Technology and Design, Singapore (e-mail: zihan\_chen@mymail.sutd.edu.sg, tonyquek@sutd.edu.sg).}
% \thanks{G.~Geraci is with the Department of Small Cells Research, Nokia Bell Labs, Dublin, Republic of Ireland (e-mail: dr.giovanni.geraci@gmail.com).}
% \thanks{Y.~Zhong is with the School of Electronic Information and Communications, Huazhong University of Science and Technology, Wuhan, P.R. China (email: yzhong@hust.edu.cn).}
%\thanks{H.~H.~Yang, Y.~Fu, and T.~Q.~S.~Quek are with the Information System Technology and Design Pillar, Singapore University of Technology and Design (e-mail: \{howard\_yang, yaru\_fu, tonyquek\}@sutd.edu.sg).
%
%J.~W.~Lin is with the College of Electrical and Computer Engineering, National Chiao Tung University, Hsinchu, Taiwan (e-mail: fiona2919.cm06g@nctu.edu.tw).
%
% \thanks{R. Srikant is with the Department of Electrical and Computer Engineering, University of Illinois at Urbana-Champaign, Champaign, IL 61820, USA (email: rsrikant@illinois.edu).}

% \thanks{H.~V.~Poor is with the Department of Electrical and Computer Engineering, Princeton University, Princeton, NJ 08544 USA (e-mail: poor@princeton.edu).}
%}
}
\maketitle
\acresetall
\thispagestyle{empty}
%%---------------------------------------------------------------------------%
%%                           abstract and key words                          %
%%---------------------------------------------------------------------------%
\begin{abstract}
% Render unto Caesar, large-scale federated learning (FL) can be achieved in a wireless network without an edge server. 
% We demonstrate that simple amplitude modulation and match filters can play similar role as the edge server in federated learning (FL), and hence large-scale FL can be achieved in a wireless network without an edge server. 
% We demonstrate that the function of an edge server in federated learning (FL) can be realized by merely analog transmissions and match filtering. 
We demonstrate that merely analog transmissions and match filtering can realize the function of an edge server in federated learning (FL).
Therefore, a network with massively distributed user equipments (UEs) can achieve large-scale FL without an edge server. 
% and hence realize FL in a network with massively distributed user equipments (UEs). 
We also develop a training algorithm that allows UEs to continuously perform local computing without being interrupted by the global parameter uploading, which exploits the full potential of UEs' processing power.
We derive convergence rates for the proposed schemes to quantify their training efficiency. 
The analyses reveal that when the interference obeys a Gaussian distribution, the proposed algorithm retrieves the convergence rate of a server-based FL. 
But if the interference distribution is heavy-tailed, then the heavier the tail, the slower the algorithm converges.  
Nonetheless, the system run time can be largely reduced by enabling computation in parallel with communication, whereas the gain is particularly pronounced when communication latency is high.
These findings are corroborated via excessive simulations.
\end{abstract}
\acresetall

%%---------------------------------------------------------------------------%
%%                           Sec: Introduction                               %
%%---------------------------------------------------------------------------%
\section{Introduction}\label{sec:intro}
% Deploying federated learning (FL) in mobile edge networks has been conceived as a core functionality of next-generation wireless systems. 
\subsection{Motivation}
A federated learning (FL) system \cite{MaMMooRam:17AISTATS,LiSahTal:20SPM,ParSamBen:19} generally consists of an edge server and a group of user equipments (UEs).
The entities collaboratively optimize a common loss function. 
The training process constitutes three steps: ($a$) each UE conducts local training and uploads the intermediate parameters, e.g., the gradients, to the server, ($b$) the server aggregates the gradients to improve the global model, and ($c$) the server broadcasts the global parameter back to the UEs for another round of local computing. 
This procedure repeats until the model converges. 
However, edge services are yet widely available in wireless networks as deploying such computing resources on the access points (APs) is costly to the operators. 
And even if possible, using the precious edge computing unit to perform global model improvement -- which are simply additions and/or multiplications -- results in significant resource underuse. 
That leads to a natural question:\\[-2.5mm] \newline
\textbf{Question-1:} \textit{Can we run FL in a wireless network without an edge server while maintaining scalability and efficiency?}\\[-2.5mm]

A possible strategy is to change the network topology from a star connection into a decentralized one \cite{LiCenChe:20AISTATS,ChePooSaa:20CMAG}. 
In this fashion, every UE only exchanges intermediate parameters with its geographically proximal neighbors in each communication round. 
If the network is fully connected, i.e., any pair of UEs can reach each other via finite hops, the training algorithm is able to eventually converge.
Nonetheless, such an approach bears two critical setbacks: ($i$) the communication efficiency is low because UEs' parameters can only be exchanged within local clusters in each global iteration, which results in a large number of communication rounds before the model can reach a satisfactory performance level; and ($ii$) the privacy issue is severe, as UEs may send their information to a deceitful neighbor without the authentication from a centralized entity. 
Therefore, completely decentralizing the network is not a desirable solution to the posed question.

Apart from putting the server in petty use, another disadvantage of the conventional FL algorithm is that once the local parameters are uploaded to the edge, UEs need to wait for the results before they can proceed to the next round of local computing. 
Since the UEs are obliged to freeze their local training during each global communication, where the latter can be orders of magnitudes slower than the former \cite{LanLeeZho:17}, the system's processing power is highly underutilized. 
As such, the second question arises: \\[-2.5mm] \newline
\textbf{Question-2:} \textit{Can UEs continue their local computing during global communication and use these extra calculations to reduce the system run time?}

\subsection{Main Contributions}
% This paper takes a different route to address the posed question. 
In light of the above challenges, we propose a new architecture, as well as the model training algorithm, that ($a$) attains a similar convergence rate of FL under the master-slave framework but without the help of an edge server and ($b$) allows local computations to be executed in parallel with global communication, therefore enhance the system's tolerance to high network latency. 
The main contributions of the present paper are summarized as follows:
\begin{itemize}
    \item We develop a distributed learning paradigm that in each communication round, allows all the UEs to simultaneously upload and aggregate their local parameters at the AP without utilizing an edge server, and later use the global model to rectify and improve the local results. This is accomplished through analog gradient aggregation \cite{SerCoh:20TSP} and replacing the locally accumulated gradients with the globally averaged ones. 
    \item We derive the convergence rate for the proposed training algorithm. The result reveals that the convergence rate is primarily dominated by the level of heavy tailedness in the interference's statistical distribution. Specifically, if the interference obeys a Gaussian distribution, the proposed algorithm retrieves the convergence rate of a conventional server-based FL. When the interference distribution is heavy-tailed, then the heavier the tail, the slower the algorithm converges.  
    \item We improve the developed algorithm by enabling UEs to continue their local computing in concurrence with the global parameter updating. We also derive the convergence rate for the new scheme. The analysis shows that the proposed method is able to reduce the system run time, and the gain is particularly pronounced in the presence of high communication latency. 
    \item We carry out extensive simulations on the MNIST and CIFAR-10 data set to examine the algorithm under different system parameters. The experiments validate that SFWFL achieves a similar, or even outperforms the convergence rate of a server-based FL, if the interference follows a Gaussian distribution. It also confirms that the convergence performance of SFWFL is sensitive to the heavy tailedness of interference distribution, where the convergence rate deteriorates quickly as the tail index decreases. Yet, as opposed to conventional FL, under the SFWFL framework, an increase in the number of UEs is instrumental in accelerating the convergence. And the system run time is shown to be drastically reduced via pipelining computing with communication. 
\end{itemize}

\begin{figure*}[t!]
  \centering

  \subfigure[\label{fig:1a}]{\includegraphics[width=0.95\columnwidth]{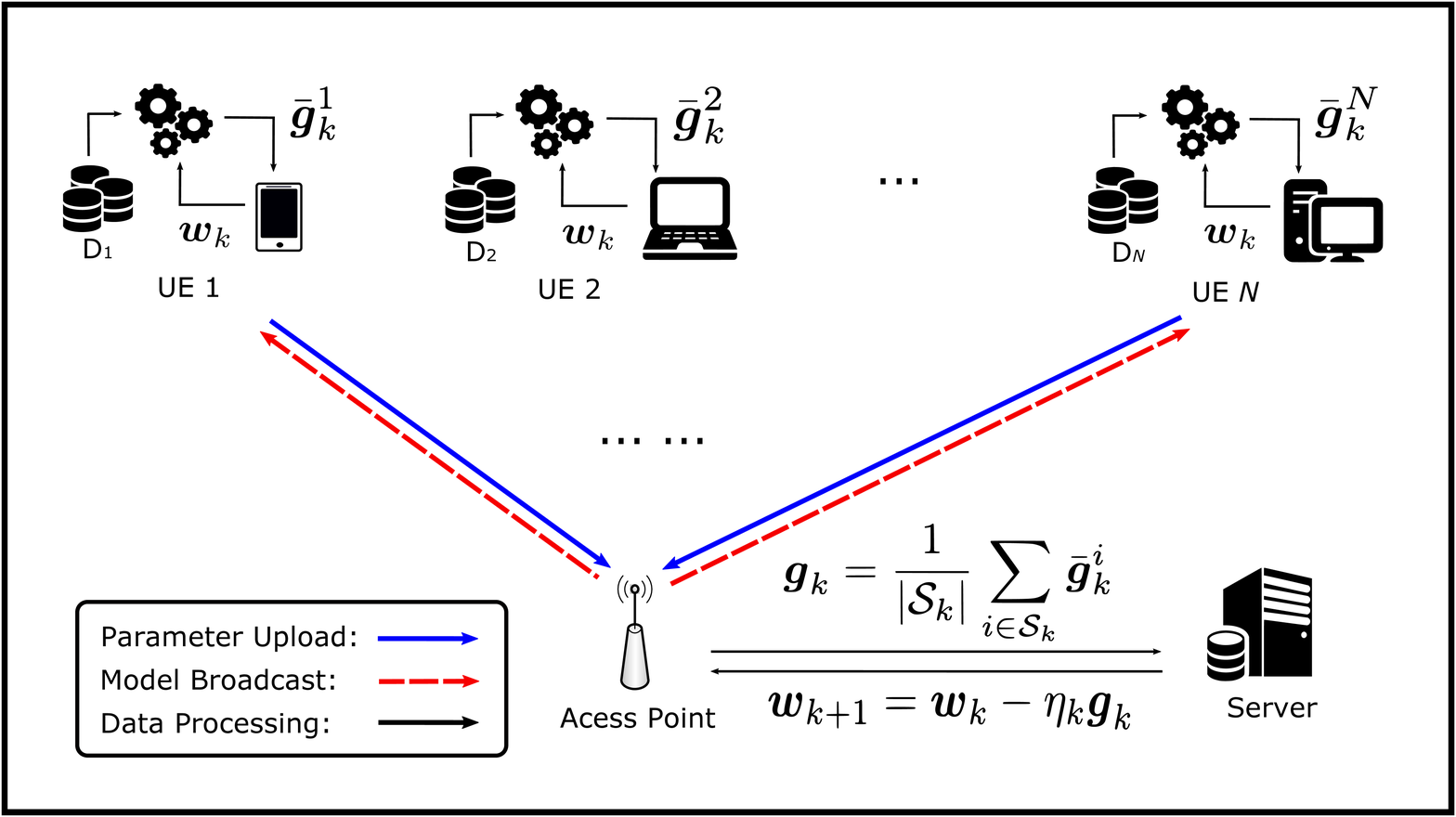}} ~~~
  \subfigure[\label{fig:1b}]{\includegraphics[width=0.95\columnwidth]{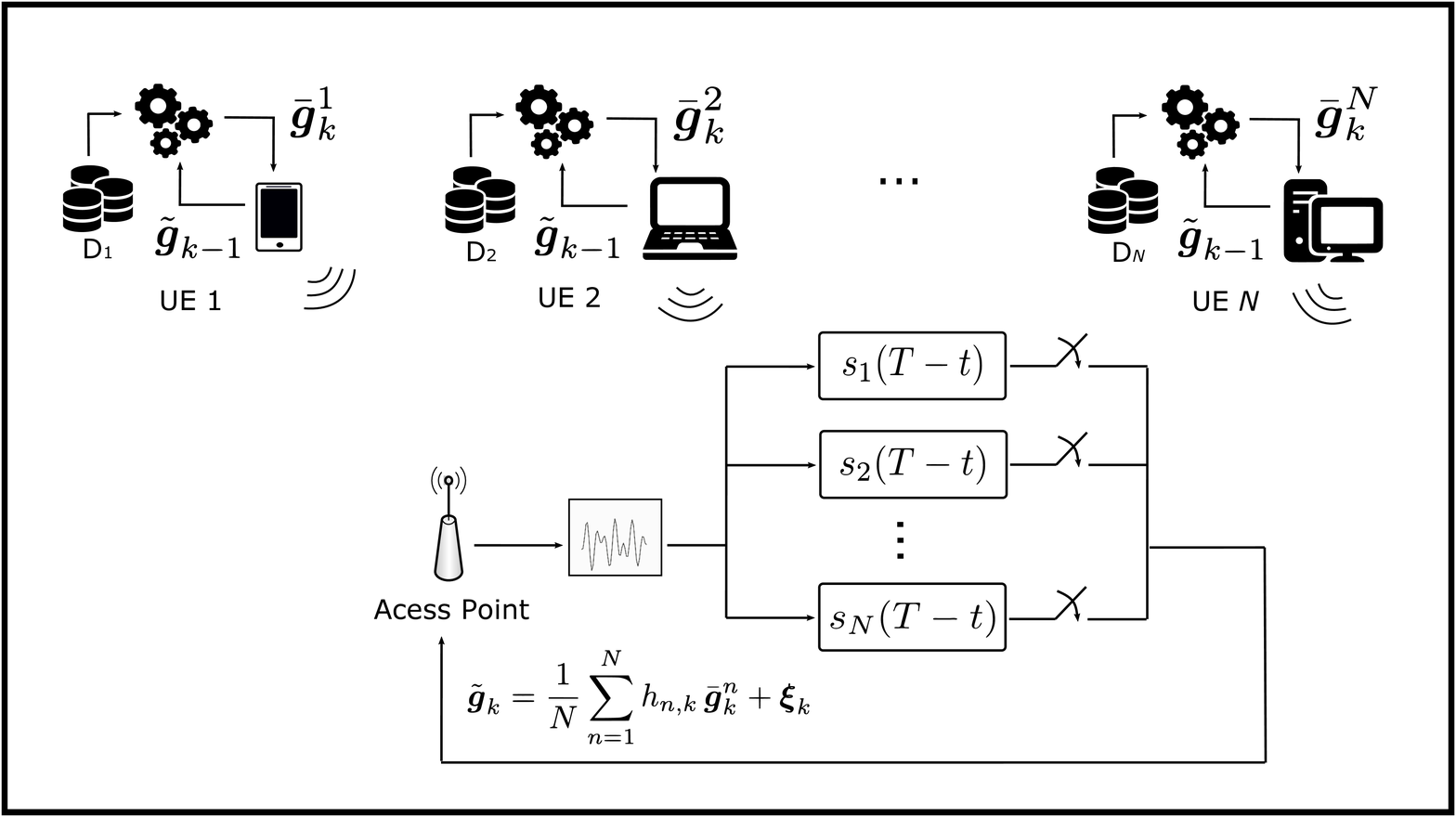}}
  \caption{ Examples of ($a$) server-based and ($b$) server free FL systems. In ($a$), the edge server aggregates the gradients from a portion of the associated UEs to improve the global model. In ($b$), all the UEs concurrently send analog functions of their gradients to the AP, and the AP passes the received signal to a bank of match filters to obtain the automatically aggregated (but noisy) gradients. }
  \label{fig:FL_Systems_comparison}
\end{figure*}

\subsection{Outline}
The remainder of this paper is organized as follows. 
We survey the related works in Section~II.
In Section~III, we introduce the system model.
We present the design and analysis of a server-free FL paradigm in Section~IV.
We develop an enhanced version of the training algorithm in Section~V, that allows UEs to execute local computations in parallel with global communications.
% In Section IV, we analyze the generalization error of the analog over-the-air GD algorithm.
Then, we show the simulation results in Section VI to validate the analyses and obtain design insights.
We conclude the paper in Section VII.

In this paper, we use bold lower case letters to denote column vectors. For any vector $\boldsymbol{w} \in \mathbb{R}^d$, we use $\Vert \boldsymbol{w} \Vert$ and $\boldsymbol{w}^{\mathsf{T}}$ to denote the $L$-2 norm and the transpose of a column vector, respectively. The main notations used throughout the paper are summarized in Table~I.

\section{Related Works}
The design and analysis of this work stem from two prior arts: \textit{Analog gradient descent} and \textit{delayed gradient averaging}. 
In the following, we elaborate on these two aspects' related works. 

\subsubsection{Analog gradient descent} 
This method capitalizes on the superposition property of electromagnetic waves for fast and scalable FL tasks \cite{ZhuWanHua:19TWC,GuoLiuLau:20,SerCoh:20TSP,YanCheQue:21JSTSP}: 
Specifically, during each global iteration, the edge server sends the global parameter to all the UEs. 
After receiving the global parameter, each UE conducts a round of local computing and, once finished, transmits an analog function of its gradient using a set of common shaping waveforms, one for each element in the gradient vector. 
The edge server receives a superposition of the analog transmitted signals, representing a distorted version of the global gradient. 
The server then updates the global model and feedbacks the update to all the UEs. 
This procedure repeats for a sufficient number of rounds until the training converges -- the convergence is guaranteed if the loss function has nice structures (i.e., strong convexity and smoothness), even if the aggregated parameters are severely jeopardized by channel fading and interference noise \cite{YanCheQue:21JSTSP}.
The main advantage of analog gradient descent is that the bandwidth requirement does not depend on the number of UEs. As a result, the system not only scales easily but also attains significant energy saving \cite{SerCoh:20TSP}. 
Moreover, the induced interference noise can be harnessed for accelerating convergence \cite{ZhaWanLi:22TWC}, enhancing privacy \cite{ElgParIss:21},  efficient sampling \cite{LiuSim:21}, or improving generalization \cite{YanCheQue:21JSTSP}. 
In addition to these benefits, the present paper unveils another blessing from the analog gradient descent, that using this method, we can get rid of the edge server -- 
as the old saying goes, ``Render unto Caesar the things which are Caesar's, and unto God the things that are God's.''

\subsubsection{Delayed gradient averaging}
On a separate track, delayed gradient averaging \cite{ZhuLinLu:21} is devised by recognizing that the gradient averaging in FL can be postponed to a future iteration without violating the federated computing paradigm.
Under delayed gradient averaging, the UEs send their parameters to each other at the end of each computing round and immediately start the next round of local training. 
The averaging step is belated to a later iteration when the aggregated result is received, upon which a gradient correction term is adopted to compensate for the staleness. 
In this manner, the communication can be pipelined with computation, hence endowing the system with a high tolerance to communication latency. 
However, \cite{ZhuLinLu:21} requires each UE to pass its parameter to every other UE for gradient aggregation, which incurs hefty communication overhead, especially when the network grows in size.
Even by adopting a server at the edge to take over the aggregation task, communication efficiency remains a bottleneck for the scheme. 
Toward this end, we incorporate analog gradient descent to circumvent the communication bottleneck of delayed gradient averaging, and show that such a marriage yields very fruitful outcomes.

% Words: gear toward, formidable challenge, make it cumbersome, recurrently interacting with, jeopardize, this investigation yields any fruit, capitalize on, ominous, manuver; Render unto Caesar the things which are Caesar's, and unto God the things that are God's; cater to; as opposed to; acquire; fidelity; As we shall see momentarily; to remedy the; in which our hopes for the future can nest and grow; at the expense of; for notational simplicity, we should henceforth assume that; succinctly; intuitively, when drawing upon ideas from centralized learning to proposed communication efficient FL methods;  

 \begin{table}
 \caption{ Notation Summary } \label{table:notation}
 \begin{center}
 \renewcommand{\arraystretch}{1.3}
 \begin{tabular}{c  p{ 5.5cm } }
 \hline
  {\bf Notation} & {\hspace{2.5cm}}{\bf Definition}
 \\
 %\midrule
 \hline
 $N$; $\mathbf{s}(t)$ & Number of UEs in the network; a set of orthonormal waveforms \\
 $M$; $D$ & Number of SGD iterations in one local computing round; number of local computing rounds in a global communication round \\
 $f( \boldsymbol{w} )$; $\nabla f( \boldsymbol{w} )$ & Global loss function; and its gradient \\
 $f_n( \boldsymbol{w} )$; $\nabla f_n( \boldsymbol{w} )$ & Local loss function of UE $n$; and its gradient \\
 $x^{n}_k(t)$; $y_k(t)$ & Analog signal sent out by UE $n$ in the $k$-th communication round; analog signal received by the AP in the $k$-th communication round \\
 $P_{n}$; $h_{n,k}$ & Transmit power of UE $n$; channel fading experience by UE $n$ \\
 $\boldsymbol{g}_k$; $\boldsymbol{\xi}_k$ & Noisy gradient received at the AP; electromagnetic interference that follows $\alpha$-stable distribution \\
 $\eta_k$ & Learning rate of the algorithm \\
 $\alpha$ & Tail index of the heavy-tailed interference \\
 $\boldsymbol{w}^{\langle \alpha \rangle}$; $\Vert \boldsymbol{w} \Vert_\alpha$  & Signed power $\alpha$ of a vector $\boldsymbol{w}$; $\alpha$-norm of a vector $\boldsymbol{w}$ \\
 \hline
 \end{tabular}
 \end{center}\vspace{-0.63cm}
 \end{table}%

%%---------------------------------------------------------------------------%
%%                           Sec: System Model                               %
%%---------------------------------------------------------------------------%
\section{System Model}

% \subsection{Setting}
We consider a wireless network consisting of one AP and $N$ UEs, as depicted in Fig.~1($b$). 
Each UE $n$ holds a loss function $f_n: \mathbb{R}^d \rightarrow \mathbb{R}$ that is constructed based on its local dataset. 
The goal of all the UEs is to jointly minimize a  global objective function. More formally, they need to cooperatively find a vector $\boldsymbol{w} \in \mathbb{R}^d$ that satisfies the following:
\begin{align} \label{equ:ObjFunc}
\underset{{ \boldsymbol{w} \in \mathbb{R}^d }}{ \mathrm{min} } ~~ f(\boldsymbol{w}) = \frac{1}{N} \sum_{ n=1 }^N f_n( \boldsymbol{w} ).
\end{align}
The solution to \eqref{equ:ObjFunc} is commonly known as the empirical risk minimizer, denoted by 
\begin{align}
\boldsymbol{w}^* = \arg\min_{ \boldsymbol{w} \in \mathbb{R}^d } f(\boldsymbol{w}).
\end{align}

In order to obtain the minimizer, the UEs need to conduct local training and periodically exchange the parameters for a global update. 
Because the AP is not equipped with a computing unit, conventional FL training schemes that rely on an edge server to perform the intermediate global aggregation and model improvement seem inapplicable in this context.
That said, we will show in the sequel that by adopting analog over-the-air computing \cite{NazGas:07IT}, one can devise an FL-like model training method that is communication efficient, highly scalable, and has the same convergence rate as the paradigms that have an edge server. 

\begin{algorithm}[t!]
\caption{ Server Free Wireless Federated Learning }
\begin{algorithmic}[1] \label{Alg:Gen_FL}
\STATE \textbf{Parameters:} $M$ = number of steps per local computing, $\eta_k$ = learning rate for the $k$-th round stochastic gradient descent.
\STATE \textbf{Initialize:} Each agent sets $\boldsymbol{w}^n_{1,1}= \boldsymbol{w}_0 \in \mathbb{R}^d$ where $\boldsymbol{w}^0$ is a randomly generated vector and $n \in \{ 1, \cdots, N \}$.
\FOR { $k = 1, \cdots, K$ }
% \STATE The AP computes the gradient of the objective function $\nabla P(\mathbf{w}^t) = \sum_{i=1}^n \nabla \ell_i( \mathbf{w}^t ) / n + \xi \nabla r(\mathbf{w}^t)$, and send parameters $\mathbf{w}^t$ and $\nabla P(\mathbf{w}^t)$ to all the UEs
\FOR { each UE $n$ in parallel }
\STATE Set the initial local parameter as follows: 
\begin{align} \label{equ:LocIntlz}
\boldsymbol{w}_{k,1}^n &= \boldsymbol{w}_{k-1, M+1}^n
\nonumber\\
&= \boldsymbol{w}_{k-1, 1}^n - \eta_{ k -1 } \left( \tilde{ \boldsymbol{g} }_{k-1} - \bar{\boldsymbol{g}}^n_{k-1} \right) 
\end{align}
in which $\boldsymbol{w}_{i, j}^n$ denotes the $j$-th iteration at round $i$, $\tilde{ \boldsymbol{g} }_{k-1}$ is the parameters received from the AP, and $\bar{\boldsymbol{g}}^n_{k-1}$ is the locally aggregated gradient. 
\FOR { $i$ = 1 to $M$ }
\STATE Sample $\gamma_i \in \mathcal{D}_k$ uniformly at random, and update the local parameter $\boldsymbol{w}^n_{k,i}$ as follows
\begin{align} \label{equ:LocGraDscnt}
\boldsymbol{w}_{k, i+1}^n = \boldsymbol{w}_{k,i}^n - \eta_k \nabla f_n(\boldsymbol{w}_{k,i}^n; \gamma_i)
\end{align}
\ENDFOR
\STATE Compute the aggregated local gradients at round $k$ as $\bar{\boldsymbol{g}}^n_{k} = \sum_{ i=1 }^M \nabla f_n(\boldsymbol{w}_{k,i}^n; \gamma_i)$, modulate $\bar{\boldsymbol{g}}^n_{k}$ onto a set of orthonormal waveforms and simultaneously send out to the AP.
\ENDFOR
\STATE The AP passes the received signal to a bank of matched filters and arrives at the following
\begin{align} \label{equ:AnlgGrdnt}
\tilde{\boldsymbol{g}}_k = \frac{1}{N} \sum_{ n=1 }^N h_{ n,k } \, \bar{ \boldsymbol{g} }^n_k + \boldsymbol{\xi}_k 
\end{align}
where $h_{ n,k }$ is the channel fading experienced by UE $n$ and $\boldsymbol{\xi}_k$ is the electromagnetic interference. The AP feeds back $\tilde{\boldsymbol{g}}_k$ to all UEs in a broadcast manner.
\ENDFOR
\STATE \textbf{Output:} $\{ \boldsymbol{w}^n_K \}_{n=1}^N$
\end{algorithmic}
\end{algorithm}

%---------------------------------------------------------------------------%
%                           Sec: SFWFL - Vanilla                            %
%---------------------------------------------------------------------------%

\section{Server Free Federated Model Training: Vanilla Version}
In this section, we detail the design and analysis of a model training paradigm that achieves similar performance to FL without the help of an edge server.
Owing to such a salient feature, we coin this scheme as \textit{Server Free Wireless Federated Learning (SFWFL)}. 
We summarize the general procedures of SFWFL in Algorithm~\ref{Alg:Gen_FL} and elaborate on the major components below. 

\subsection{Design}
Similar to the conventional FL, SFWFL requires local trainings at the UEs, global communications of intermediate parameters, and feedback of the aggregated results. 

\subsubsection{Local Training}
Before the training commences, all UEs negotiate amongst each other on an initial parameter $\boldsymbol{w}_0$ that is randomly generated. 
Then, every UE conducts $M$ steps of SGD iteration based on its own dataset and updates the locally aggregated gradient to the AP.
The AP (automatically) aggregates the UEs' gradients by means of analog over-the-air computing--which will be elucidated soon--and feeds back the resultant parameters to all the UEs. 
Upon receiving the globally aggregated gradient, every UE replaces the locally aggregated gradient by this global parameter, as per \eqref{equ:LocIntlz}, and proceeds to the next round of local computing in accordance with \eqref{equ:LocGraDscnt}. 

It is important to note that by replacing the local gradients with the global one, the UEs' model parameters $\{ \boldsymbol{w}^n_k \}_{n=1}^N$ are aligned at the beginning of each local computing stage. 
As such, if the model training converges, every UE will have its parameters approach the same value. 

\subsubsection{Global Communication}
During the $k$-th round of global communication, UE $n$ gathers the stochastic gradients calculated in the current computing round as $\bar{\boldsymbol{g}}^n_{k} = \sum_{ i=1 }^M \nabla f_n(\boldsymbol{w}_{k,i}^n; \gamma_i)$, and constructs the following analog signal:
\begin{align} \label{equ:AnagMod}
x^n_k(t) = \langle \, \mathbf{s}(t), \, \bar{\boldsymbol{g}}^n_{k} \, \rangle
\end{align}
where $\langle \cdot, \cdot \rangle$ denotes the inner product between two vectors and $\mathbf{s}(t) = (s_1(t), s_2(t), ..., s_d(t))$, $0 < t <T$, is a set of orthonormal baseband waveforms that satisfies:
\begin{align}
&\int_0^T s^2_i(t) dt = 1,~~~ i = 1, 2, ..., d \\
&\int_0^T s_i(t) s_j(t) = 0, ~~~\text{if}~ i \neq j.
\end{align}
In essence, operation \eqref{equ:AnagMod} modulates the amplitude of $s_i(t)$ according to the $i$-th entry of $\bar{\boldsymbol{g}}^n_{k}$ and superpositions the signals into an analog waveform.
% According to \eqref{equ:AnagMod}, the signal $x_n(t)$ is essentially a superposition of the analog waveforms whereas the magnitude of $s_i(t)$ equals to the $i$-th element of $\nabla f_n( \boldsymbol{w}_k  )$.\footnote{Note that the magnitude of the waveforms can also be set at the quantized values of the gradients to reduce implementation complexity.}
Once the transmit waveforms $\{x^n_k(t)\}_{n=1}^N$ have been assembled, the UEs send them out concurrently into the spectrum. 
We consider the UEs employ power control to compensate for the large-scale path loss while the instantaneous channel fading are unknown. 

Notably, since the waveform basis are independent to the number of UEs, this architecture is highly scalable. In other words, \textit{all the UEs can participate in every round of local training and global communication regardless of how many UEs are there in the network.} 

\subsubsection{Gradient Aggregation}
The analog signals go through wireless medium and accumulated at the AP's RF front end. 
The received waveform can be expressed as follows:{\footnote{In this paper, we consider the waveforms of different UEs are synchronized. Note that the issue of signal misalignment can be addressed via \cite{ShaGunLie:21TWC}. }} 
\begin{align}
y_k(t) = \sum_{ n=1 }^N h_{n, t} x^n_k(t) + \xi (t)
\end{align}
where $h_{n, t}$ is the channel fading experienced by UE $n$ and $\xi(t)$ stands for the interference. 
Without loss of generality, we assume the channel fading is independent and identically distributed (i.i.d.) across the agents and communication rounds, with a unit mean and variance $\sigma^2$. 
Furthermore, we consider $\xi(t)$ follows a symmetric $\alpha$-stable distribution \cite{ClaPedRod:20}, which is widely used in characterizing interference's statistical property in wireless networks \cite{Mid:77,WinPin:09,YanPet:03}.

The AP passes the analog signal $y_k(t)$ to a bank of matched filters, where each branch is tuned to one element of the waveform basis, and outputs the vector in \eqref{equ:AnlgGrdnt}, where $\boldsymbol{\xi}_k$ is a $d$-dimensional random vector with each entry being i.i.d. and following an $\alpha$-stable distribution. 
The AP then broadcasts $\tilde{ \boldsymbol{g} }_k$ back to all the UEs. 
Owing to the high transmit power of the AP, we assume the global parameters can be received without error by all the UEs.
Then, the UEs move to step-1) and launch a new round of local computing.

The most remarkable feature of this model training paradigm is that it does not require an edge server to conduct global aggregation and/or model improvement.
Instead, the AP exploits the superposition property of wireless signals to achieve fast gradient aggregation through a bank of match filters. 
At the UEs side, they replace the locally aggregated gradient with the global one at the beginning of each local computing round to align the model parameters. 
As will be shown next, the training algorithm converges albeit the global gradients are highly distorted. 
Apart from server free, it is noteworthy that since the UEs do not need to compensate for the channel fading, they can transmit at a relatively constant power level to save the hardware cost. 
Additionally, the random perturbation from fading and interference provides inherent privacy protection to the UEs' gradient information \cite{ElgParIss:21}.

\subsection{Analysis}
In this part, we derive the convergence rate to quantify the training efficiency of SFWFL. 
\subsubsection{Preliminary Assumptions}
To facilitate the analysis, we make the following assumptions.
\begin{assumption}
\textit{The objective functions $f_n: \mathbb{R}^d \rightarrow \mathbb{R}$ are $\mu$-strongly convex, i.e., for any $\boldsymbol{w}, \boldsymbol{v} \in \mathbb{R}^d$ it is satisfied: 
    \begin{align}
    f_n( \boldsymbol{w} ) \geq f_n( \boldsymbol{v} ) + \langle \nabla f_n( \boldsymbol{v} ), \boldsymbol{w} - \boldsymbol{v} \rangle + \frac{\mu}{2} \Vert \boldsymbol{w} - \boldsymbol{v} \Vert^2.
    \end{align}    
}
\end{assumption}

\begin{assumption}
\textit{The objective functions $f_n: \mathbb{R}^d \rightarrow \mathbb{R}$ are $\lambda$-smooth, i.e., for any $\boldsymbol{w}, \boldsymbol{v} \in \mathbb{R}^d$ it is satisfied: 
    \begin{align}
    f_n( \boldsymbol{w} ) \leq f_n( \boldsymbol{v} ) + \langle \nabla f_n( \boldsymbol{v} ), \boldsymbol{w} - \boldsymbol{v} \rangle + \frac{\lambda}{2} \Vert \boldsymbol{w} - \boldsymbol{v} \Vert^2.
    \end{align}    
}
\end{assumption}

\begin{assumption}
\textit{The stochastic gradients are unbiased and have bounded second moments, i.e., there exists a constant $G$ such that the following holds:
    \begin{align}
     \mathbb{E}\left[ \Vert \nabla f_n(\boldsymbol{w}_{k,i}^n; \gamma_i) \Vert^2 \right] \leq G^2.
    \end{align}    
}
\end{assumption}

Because the interference follows an $\alpha$-stable distribution, which has finite moments only up to the order $\alpha$, the variance of the globally aggregated gradient in \eqref{equ:AnlgGrdnt} may be unbounded. 
As such, conventional approaches that rely on the existence of second moments cannot be directly applied. 
In order to establish a universally applicable convergence analysis, we opt for the $\alpha$-norm as an alternative. 
Based on this metric, we introduce two concepts, i.e., the \textit{signed power} and \textit{$\alpha$-positive definite matrix} \cite{WanGurZhu:21}, in below. 

\begin{definition}
\textit{For a vector $\boldsymbol{w} = (w_1, ..., w_d)^{\mathsf{T}} \in \mathbb{R}^d$, we define its signed power as follows
\begin{align}
\boldsymbol{w}^{\langle \alpha \rangle } = \left( \mathrm{sgn}(w_1) \vert w_1 \vert^\alpha, ...,  \mathrm{sgn}(w_d) \vert w_d \vert^\alpha \right)^{\mathsf{T}}
\end{align}
where $\mathrm{sgn}(x) \in \{ -1, +1 \}$ takes the sign of the variable $x$. }
\end{definition}

\begin{definition}
\textit{A symmetric matrix $\boldsymbol{Q}$ is said to be $\alpha$-positive definite if $\langle \boldsymbol{v}, \boldsymbol{Q} \boldsymbol{v}^{\langle \alpha - 1 \rangle} \rangle > 0$ for all $\boldsymbol{v} \in \mathbb{R}^d$ with $\Vert \boldsymbol{v} \Vert_\alpha > 1$.    
}
\end{definition}

Armed with the above definitions, we make another assumption as follows.
\begin{assumption}
\textit{For any given vector $\boldsymbol{w} \in \mathbb{R}^d$, the Hessian matrix of $f(\boldsymbol{w})$, i.e., $\nabla^2 f(\boldsymbol{w})$, is $\alpha$-positive definite. }
\end{assumption}

Furthermore, since each element of $\boldsymbol{\xi}_k$ has a finite $\alpha$ moment, we consider the $\alpha$ moment of $\boldsymbol{\xi}_k$ is upper bounded by a constant $C$, i.e., $\mathbb{E}[ \Vert \boldsymbol{\xi}_k \Vert_\alpha^\alpha ] \leq C$.

\subsubsection{Convergence Rate of SFWFL}
We lay out two technical lemmas that we would use extensively in the derivation. 
\begin{lemma}
\textit{Given $\alpha \in [1, 2]$, for any $\boldsymbol{w}, \boldsymbol{v} \in \mathbb{R}^d$, the following holds:
\begin{align}
\Vert \boldsymbol{w} + \boldsymbol{v} \Vert_\alpha^\alpha \leq \Vert \boldsymbol{w} \Vert_\alpha^\alpha + \alpha \langle \boldsymbol{w}^{\langle \alpha - 1 \rangle }, \boldsymbol{v} \rangle + 4 \Vert \boldsymbol{v} \Vert_\alpha^\alpha.
\end{align}
}
\end{lemma}
\begin{IEEEproof}
Please refer to \cite{Kar:69}.
\end{IEEEproof}

\begin{lemma}
\textit{Let $\boldsymbol{Q}$ be an $\alpha$-positive definite matrix, for $\alpha \in [1, 2]$, there exists $\kappa, L >0$, such that
\begin{align}
\Vert \boldsymbol{I} - k \boldsymbol{Q} \Vert_\alpha^\alpha \leq 1 - L k, \qquad \quad \forall k \in [0, \kappa)
\end{align}
 }
\end{lemma}
\begin{IEEEproof}
Please see Theorem~10 of \cite{WanGurZhu:21}. 
\end{IEEEproof}
% \begin{lemma}
% \textit{For a sequence of real numbers $\{ b_k \}$, $k \geq 1$, that satisfies:
% \begin{align}
% b_{k+1} \leq \left( 1 - \frac{c}{k} \right) b_k + \frac{c_1}{k^{p+1}}
% \end{align}
% where $c > p> 0$, $c_1 > 0$. The following relationship holds
% \begin{align}
% b_k \leq \frac{ c_1 }{ c - p } \cdot \frac{1}{k^p} + o\left( \frac{1}{k^{p+1}} + \frac{1}{k^c} \right)
% \end{align}
% where $o(\cdot)$ is the ``little o'' notation, meaning that if $f(k) = o(g(k))$ then $\forall B > 0$, there exists $k_0$ such that $f(k) \leq B g(k)$ for all $k \geq k_0$.
% }
% \end{lemma}
% \begin{IEEEproof}
% Please see Lemma~1 of \cite{Chung:54}. 
% \end{IEEEproof}

% We are now ready to present the first theoretical finding of this paper. 

Since the UEs' model parameters are aligned at the beginning of each local computing round, i.e., $\boldsymbol{w}^1_{k,1} = \boldsymbol{w}^2_{k,1} = \cdots = \boldsymbol{w}^N_{k,1}$, we denote such a quantity as $\boldsymbol{w}_k$ and present the first theoretical finding below. 

\begin{theorem} \label{thm:ConvAnals}
\textit{Under the employed wireless system, if the learning rate is set as $\eta_k = \theta/k$ where $\theta > \frac{ \alpha - 1 }{ M L }$, then Algorithm-1 converges as:
    % \begin{align} \label{equ:ConvRt_OAML}
    % \mathbb{E}\big[ \Vert \boldsymbol{w}_k - \boldsymbol{w}^* \Vert_{\alpha}^\alpha \big] \!\leq \! \frac{ 4 \theta^\alpha \Big( C \!+ d^{ 1 - \frac{1}{\alpha} } G^\alpha M^\alpha \big( \lambda^\alpha M^\alpha \!+\! \frac{ \sigma^\alpha   }{ N^{\alpha/2} } \big)  \Big) }{ \big( \mu M L - \alpha + 1 \big) \,  k^{\alpha - 1} }.
    % \end{align}
    \begin{align} \label{equ:ConvRt_OAML}
    &\mathbb{E}\big[ \Vert \boldsymbol{w}_k - \boldsymbol{w}^* \Vert_{\alpha}^\alpha \big] 
    \nonumber\\
    &\leq \! \frac{ 4 \theta^\alpha \Big( C + d^{ 1 - \frac{1}{\alpha} } G^\alpha M^\alpha \big( \lambda^\alpha M^\alpha + \frac{ \sigma^\alpha   }{ N^{\alpha/2} } \big)  \Big) }{ \mu M L - \alpha + 1  } \cdot \frac{1}{ k^{\alpha - 1} }.
    \end{align}    
}
\end{theorem}
\begin{IEEEproof}
See Appendix~\ref{Apndx:ConvAnals_proof}. 
\end{IEEEproof}

We highlight a few important observations from this result. 

\remark{
    \textit{If the interference follows a Gaussian distribution, i.e., $\alpha = 2$, the SFWFL converges in the order of $\mathcal{O}( \frac{1}{ k } )$, which is identical to those run under federated edge learning systems \cite{SerCoh:20TSP}. As such, the proposed model training algorithm can attain the same efficacy of conventional edge learning without the requirement of a server. }
}

\remark{
    \textit{The interference's tail index, $\alpha$, plays an essential role in the convergence rate. Specifically, the smaller the value of $\alpha$, the heavier the tail in the interference distribution. And that leads to slower convergence of the learning algorithm. }
}

\remark{
    \textit{Advanced signal processing techniques can be adopted for denoising the aggregated gradient, which results in reducing the variance of channel fading, $\sigma$, and accelerating the model training process. The effect is reflected in the multiplier of the convergence rate. }
}

\remark{
    \textit{An increase in the number of UEs, $N$, can mitigate the impact of channel fading and speedup the convergence rate. Therefore, scaling up the system is, in fact, beneficial to the model training under SFWFL.  }
}

\begin{algorithm}[t!]
\caption{ Zero-Wait SFWFL }
\begin{algorithmic}[1] \label{Alg:ZW_SFWFL}
\STATE \textbf{Parameters:} $M$ = number of steps per local computing, $D$ = communication latency represented in the number of computing rounds, $\eta_k$ = learning rate for the $k$-th round stochastic gradient descent.
\STATE \textbf{Initialize:} Each agent sets $\boldsymbol{w}^n_{1,1}= \boldsymbol{w}_0 \in \mathbb{R}^d$ where $\boldsymbol{w}^0$ is a randomly generated vector and $n \in \{ 1, \cdots, N \}$.
\FOR { $k = 1, \cdots, K + KD$ }
% \STATE The AP computes the gradient of the objective function $\nabla P(\mathbf{w}^t) = \sum_{i=1}^n \nabla \ell_i( \mathbf{w}^t ) / n + \xi \nabla r(\mathbf{w}^t)$, and send parameters $\mathbf{w}^t$ and $\nabla P(\mathbf{w}^t)$ to all the UEs
\FOR { each UE $n$ in parallel }
\STATE Set the initial local parameter as follows: 
\begin{align} \label{equ:ZW_Intliz}
\boldsymbol{w}_{k,1}^n 
= \! \begin{cases} 
\boldsymbol{w}_{k-1,1}^n - \eta_{k-1} \bar{\boldsymbol{g}}^n_{k-1}, ~~~ \mbox{if } \mbox{$k$ mod $D \neq 1$}, \\
\quad\\
\boldsymbol{w}_{k-1,1}^n - \eta_{k-1} \bar{\boldsymbol{g}}^n_{k-1} \nonumber\\
\qquad - \eta_{ k -D } \left( \tilde{ \boldsymbol{g} }_{k-D} - \bar{\boldsymbol{g}}^n_{k-D} \right), ~ \mbox{otherwise.}
\end{cases}
\end{align}
where $\boldsymbol{w}_{k-1, M+1}^n$ stands for the parameters calculated in round ($k-1$), $\tilde{ \boldsymbol{g} }_{k-D}$ is the global gradient received from the AP, and $\bar{\boldsymbol{g}}^n_{k-D}$ is the corresponding locally aggregated gradient. 
\FOR { $i$ = 1 to $M$ }
\STATE Sample $\gamma_i \in \mathcal{D}_k$ uniformly at random, and update the local parameter $\boldsymbol{w}^n_{k,i}$ as per \eqref{equ:LocGraDscnt}.
%  follows
% \begin{align} \label{equ:ZW_LocGraDscnt}
% \boldsymbol{w}_{k, i+1}^n = \boldsymbol{w}_{k,i}^n - \eta_k \nabla f_n(\boldsymbol{w}_{k,i}^n; \gamma_i)
% \end{align}
\ENDFOR
\IF {$k$ mod $D = 1$}
\STATE Compute the aggregated local gradients at round $k$ as $\bar{\boldsymbol{g}}^n_{k} = \sum_{ i=1 }^M \nabla f_n(\boldsymbol{w}_{k,i}^n; \gamma_i)$, modulate $\bar{\boldsymbol{g}}^n_{k}$ onto a set of orthonormal waveforms and simultaneously send out to the AP.
\ENDIF 
\ENDFOR
\STATE Upon receiving signals from the UEs, the AP passes the received waveform to a bank of matched filters and obtains $\tilde{\boldsymbol{g}}_k$ as \eqref{equ:AnlgGrdnt}. The AP then feeds back $\tilde{\boldsymbol{g}}_k$ to all UEs in a broadcast manner.
\ENDFOR
\STATE \textbf{Output:} $\{ \boldsymbol{w}^n_{K + KD} \}_{n=1}^N$
\end{algorithmic}
\end{algorithm}

\begin{figure*}[t!]
  \centering

    {\includegraphics[width=1.98\columnwidth]{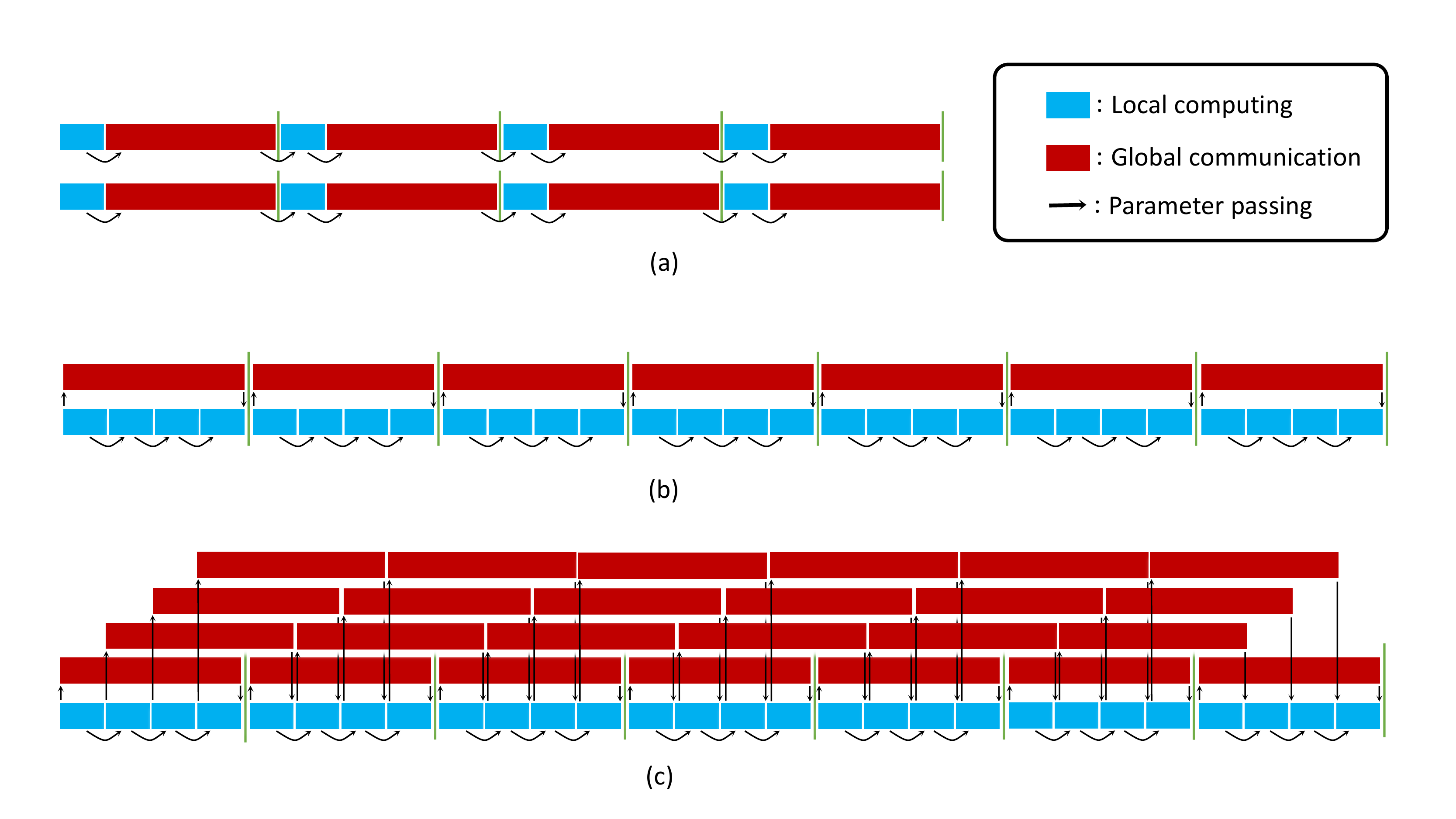}}
  \caption{ Comparison among the ($a$) compute-and-wait, ($b$) zero wait, and ($c$) aggressively updating model training schemes. In ($a$), local computing and global updating are conducted sequentially. In ($b$), local computing is executed in parallel to global communication. In ($c$), locally trained results are sent out upon finishing each round of on-device computing. }
  \label{fig:FL_commun_comparison}
\end{figure*}

%---------------------------------------------------------------------------%
%                           Sec: SFWFL - Zero Wait                          %
%---------------------------------------------------------------------------%
\section{Server Free Federated Model Training: Zero-Wait Version}
% Based on the paradigm developed in the previous section, this section presents an improvement to the learning algorithm which allows UEs to continue their local training without waiting for the global parameters. 
% As a result, the system run time can be vastly reduced.
% Due to the property that UEs no longer need to wait for the global results, we call the proposed scheme \textit{Zero Wait SFWFL}.
% More details are provided below.

This section improves the learning algorithm, allowing UEs to continue their local training without waiting for the global parameters. 
As a result, the system run time can be largely reduced. Due to the property that UEs nolonger need to wait for the global results, we call the proposed scheme \textit{Zero Wait SFWFL}. More details are provided below. 

\subsection{Design}
To simplify the notation use, let us assume every UE in the network spent the same amount of time in parameter uploading and downloading during each round of global communication. 
The communication time spreads over $D$ local computing rounds, which is in total $MD$ SGD iterations. 
In order to better illustrate the concepts, we slightly abuse the notation $k$ in this section by referring to it as the $k$-th \textit{computing} round. 
Then, as outlined in Algorithm~\ref{Alg:ZW_SFWFL}, the Zero-Wait SFWFL no longer freezes the local computation power during the communication. 
Specifically, in the first round of local computing, each UE $n$ executes $M$ SGD iterations and arrives at the following:
\begin{align}
\boldsymbol{w}^n_{1, M + 1} &= \boldsymbol{w}^n_{1, M} - \eta_1 \nabla f_n( \boldsymbol{w}^n_{1,M}; \gamma_{1,M} )
\nonumber\\
&= \cdots = \boldsymbol{w}_0 - \eta_1 \sum_{ i = 1 }^M \nabla f_n( \boldsymbol{w}^n_{1,i}; \gamma_{1,i} )
\nonumber\\
&= \boldsymbol{w}_0 - \eta_1 \, \bar{ \boldsymbol{g} }^n_1.
\end{align} 
Upon finishing the first computating round, the UEs simultaneously send their accumulated gradients $\{ \bar{\boldsymbol{g}}^n_1 \}_{ n=1 }^N$ to the AP via analog transmissions. 
Then, unlike SFWFL, the UEs do not wait for the return of the global aggregation but immediately proceed to the next round of local computation. 
By the time the globally aggregated gradients are received, UE $n$ has already performed $D$ extra rounds of local updates and its model parameter can be written as:
\begin{align}
\boldsymbol{w}^n_{1+D, 1} &= \boldsymbol{w}^n_{D, M} - \eta_D \nabla f_n( \boldsymbol{w}^n_{D,M}; \gamma_{D,M} )  
\nonumber\\
&= \boldsymbol{w}^n_{D, 1} - \eta_D \, \bar{ \boldsymbol{g} }^n_D
\nonumber\\
&= \boldsymbol{w}_0 - \eta_D \bar{ \boldsymbol{g} }^n_D - \eta_{D-1} \bar{ \boldsymbol{g} }^n_{D-1} - \cdots - \eta_1 \bar{ \boldsymbol{g} }^n_1.
\end{align}

As the aggregated gradient, $\tilde{\boldsymbol{g}}_1$, of the first round is now available, UE $n$ can substitute all the first round local gradients by the global one, which yields:
\begin{align}
\boldsymbol{w}^n_{1+D, 1} &= ( \boldsymbol{w}_0 - \eta_1 \tilde{\boldsymbol{g}}_1 ) - \eta_D \bar{ \boldsymbol{g} }^n_D - \cdots - \eta_2 \bar{ \boldsymbol{g} }^n_2
\nonumber\\
&= \boldsymbol{w}^n_{D, 1} - \eta_D \, \bar{ \boldsymbol{g} }^n_D - \eta_1 ( \tilde{\boldsymbol{g}}_1 - \bar{ \boldsymbol{g} }^n_1 ).
\end{align}

This operation serves as the cornerstone of Algorithm~2. 
By successively replacing the local gradients $\bar{ \boldsymbol{g} }^n_k$ by the globally averaged $\tilde{\boldsymbol{g}}_k$, where $k \geq 2$, the UE $n$ model parameters $\boldsymbol{w}^n_k$ approach to that in \eqref{equ:LocIntlz}. 
As such, the local models converge to the global optimality. 

A comparison between the sequential training procedure where computations are suspended during global communications and the zero wait learning scheme that pipelines the on-device computing and parameter updating is provided in Fig.~2($a$) and Fig.~2($b$). 
This figure shows that the zero wait operation moves the gradient averaging to a later stage. Since the UEs no longer need to wait for the global model, they can compute local updates restlessly. In this sense, the devices' processing powers are fully exploited, which significantly accelerates the training process.

\subsection{Analysis}
In this part, we formally demonstrate the convergence of Zero-Wait SFWFL based on the assumptions made in Section~IV-B-1). 
To facilitate the presentation, we denote the initialized model parameter of UE $n$ in computing round $k$ as $\boldsymbol{w}^n_k$, i.e., $\boldsymbol{w}^n_k = \boldsymbol{w}^n_{k,1}$.
We further denote $\bar{\boldsymbol{w}}_k$ as the averaged model parameter, given by 
\begin{align}
\bar{ \boldsymbol{w} }_k = \frac{ 1 }{ N } \sum_{ n=1 }^N \boldsymbol{w}^n_k.
\end{align}
Then, according to Algorithm~\ref{Alg:ZW_SFWFL}, the following holds
\begin{align} \label{equ:Glbl_PrmAve}
\bar{ \boldsymbol{w} }_{k+1} = \bar{ \boldsymbol{w} }_k - \frac{ \eta_k }{ N } \sum_{ n=1 }^N \bar{\boldsymbol{g}}^n_k - \eta_{ k - D } \!  \underbrace{ \left( \tilde{\boldsymbol{g}}_{k-D} - \frac{ 1 }{ N } \sum_{ n=1 }^N \bar{ \boldsymbol{g} }^n_{k-D} \right) }_{ \text{ gradient correction } }. 
\end{align}
This result indicates that the UEs' gradients are aligned from round $(k-D)$ backwards via replacing the locally accumulated gradients by the global one, which is reflected by the gradient correction in \eqref{equ:Glbl_PrmAve}.
As such, parameters of the UEs only differ on the most recent $D$ local gradients, which can be characterized formally as follows.
\begin{lemma} \label{lma:PrmtUnivBnd}
\textit{The difference between the model parameter of UE $n$ and that averaged across all UEs is uniformly bounded by the following:
\begin{align} \label{equ:GrdVar_Bnd}
\mathbb{E}\left[ \Vert \boldsymbol{w}^n_k - \bar{ \boldsymbol{w} }_k \Vert^2_2 \right] \leq 4 \eta^2_{ k - D } \, M^2 D^2 G^2.
\end{align}
}
\end{lemma}
\begin{IEEEproof}
See Appendix~\ref{Apndx:PrmtUnivBnd_proof}.
\end{IEEEproof}

From \eqref{equ:GrdVar_Bnd}, we can see that if the system adopts a diminishing learning rate, the UE's local parameters will converge to the same global average. 
In this regard, we can concentrate on the convergence performance of $\bar{ \boldsymbol{w} }_k$, which is given by the following theorem. 

\begin{theorem} \label{thm:ZW_SFWFL_ConvAnals}
\textit{Under the employed wireless system, if the learning rate is set as $\eta_k = \theta/k$ where $\theta > \frac{ \alpha - 1 }{ M L }$, then Algorithm-2 converges as:
    \begin{align} \label{equ:ConvRt_ZW_SFWFL}
    &\mathbb{E}\big[ \Vert \bar{ \boldsymbol{w}}_k - \boldsymbol{w}^* \Vert_{\alpha}^\alpha \big] 
    \nonumber\\
    &\leq \! \frac{ 4 \theta^\alpha \Big( C + \frac{ \sigma^\alpha M^\alpha G^\alpha d^{ 1 - \frac{1}{\alpha} } }{ N^{\alpha/2} } + d^{ 1 - \frac{1}{\alpha} } 2^\alpha \lambda^\alpha ( 1 \!+\! D )^\alpha G^\alpha M^{2 \alpha} \Big) }{ \big( \, \mu M L - \alpha + 1 \, \big) \cdot k^{ \alpha - 1 } } .
    \end{align}    
}
\end{theorem}
\begin{IEEEproof}
See Appendix~\ref{Apndx:ZW_SFWFL_ConvAnals_proof}.
\end{IEEEproof}
We can conclude the following observations from this result. 

\remark{
    \textit{While an increase in the communication latency, $D$, will slow down the model training, Zero-Wait SFWFL converges in the order of $\mathcal{O}( \frac{1}{ k^{ \alpha - 1 } } )$ , which is the same as those in conventional wireless FL systems. Additionally, since the communication is fully covered by local computing, the total run time can be reduced by a factor of $D$. 
    Such an improvement is particularly pronounced in mobile applications where communication delays are orders of magnitudes higher than the on-device computing \cite{LanLeeZho:17}.}
}

\remark{
    \textit{The proposed algorithm can be further accelerated by running the local SGD iterations in tandem with the momentum method \cite{YanCheQue:21JSTSP}. And one shall adequately adjust the momentum controlling factor to achieve the best training performance. }
}

\subsection{Special Case}
We can further reap the potential of the zero-wait learning method in Algorithm~\ref{Alg:ZW_SFWFL} for fast model training. 
Particularly, as depicted in Fig.~2($c$), the UEs collaboratively train the model on the basis of Zero-Wait SFWFL. 
In lieu of uploading local gradients to the AP only when global parameters are received, the UEs send their accumulated gradients upon completing each round of local training. 
As such, the UEs can always replace a local gradient by the global average in the subsequent computing rounds, making the local parameters differ from the global one in only one round of computation.
Consequently, the convergence rate can be obtained as the following.
\begin{corollary}
\textit{Under the setting in this section, if the learning rate is set as $\eta_k = \theta/k$ where $\theta > \frac{ \alpha - 1 }{ M L }$, then Algorithm-2 converges as:
    \begin{align} \label{equ:ConvRt_FPZW_SFWFL}
    &\mathbb{E}\big[ \Vert \bar{ \boldsymbol{w}}_k - \boldsymbol{w}^* \Vert_{\alpha}^\alpha \big] 
    \nonumber\\
    &\leq \! \frac{ 4 \theta^\alpha \Big( C + \frac{ \sigma^\alpha M^\alpha G^\alpha d^{ 1 - \frac{1}{\alpha} } }{ N^{\alpha/2} } + d^{ 1 - \frac{1}{\alpha} } 4^\alpha \lambda^\alpha G^\alpha M^{2 \alpha} \Big) }{ \big( \, \mu M L - \alpha + 1 \, \big) \cdot k^{ \alpha - 1 } } .
    \end{align}    
}
\end{corollary}
\begin{IEEEproof}
The result follows by substituting $D=1$ in \eqref{equ:ConvRt_ZW_SFWFL}.
\end{IEEEproof}

Following Corollary~1, we can see that the convergence rate does not depend on the communication latency $D$.
This observation implies that by aggressively uploading local parameters in each computing round, Zero-Wait SFWFL can mitigate the straggler issue of FL, at the cost of additional communication expenditure.

%---------------------------------------------------------------------------%
%                           Sec: Numerical                                  %
%---------------------------------------------------------------------------%
\section{ Simulation Results }\label{sec:NumResult}
In this section, we conduct experimental evaluations of the proposed SFWFL algorithm.
Particularly, we examine the performance of the proposed algorithm for two different tasks: ($i$) training a multi layer perceptron (MLP) on the MNIST dataset which contains the hand-written digits \cite{LeCBotBen:98} and ($ii$) learning a convolutional neural network (CNN) on the CIFAR-10 dataset \cite{KriHin:09}. 
The MLP is consisted of 2 hidden layers, each has 64 units and adopts the ReLu activations. 
We extract 60,000 data points from the MNIST dataset for training, where each UE is assigned with an independent portion that contains 600 data samples. 
For the non-IID setting on MNIST dataset, we use the 2-class configuration  \cite{MaMMooRam:17AISTATS}, in which each UE is assigned with images from  at most 2 classes.
The CIFAR-10 dataset consists of 60,000 colour images in 10 classes, with 6000 images per class. And the CNN has two convolutional layers with a combination of max pooling, followed by two fully-connected layers, then a softmax output layer. 
We extract 50,000 data points from the CIFAR-10 dataset for training, where each agent is assigned with an independent portion that contains 500 data samples.
We allocate 10,000 data points for testing. 
Furthermore, we adopt the Rayleigh fading to model the channel gain. 
Unless otherwise stated, the following parameters will be used: Tail index $\alpha = 1.6$, number of agents $N=100$, average channel gain $\mu = 1$. 
The experiments are implemented with Pytorch on Tesla P100 GPU and averaged over 3 trials.

\subsection{Numerical Results}

\begin{figure}[t!]
  \centering

  \subfigure[\label{fig:2a}]{\includegraphics[width=0.95\columnwidth]{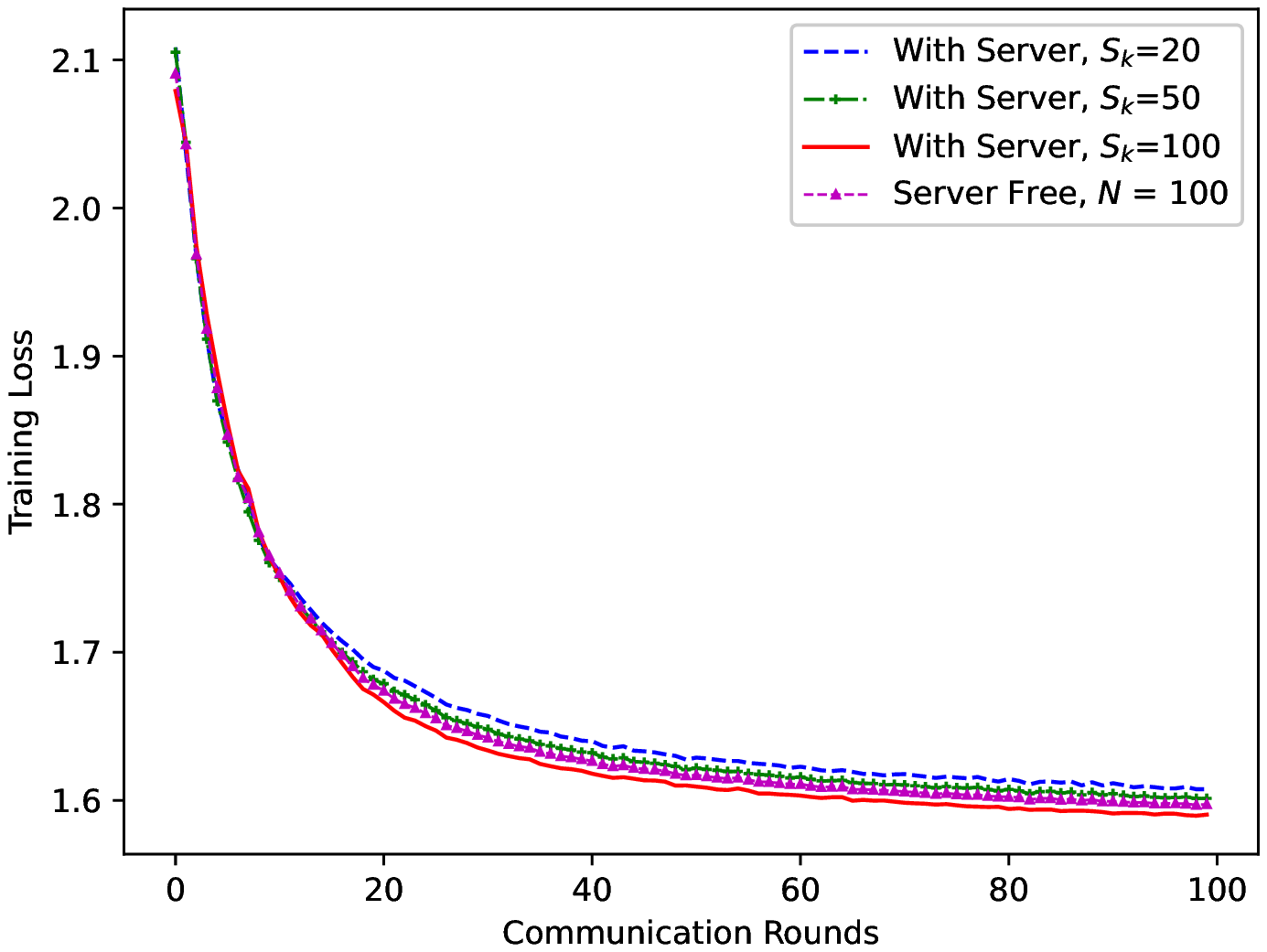}} ~
  \subfigure[\label{fig:2b}]{\includegraphics[width=0.95\columnwidth]{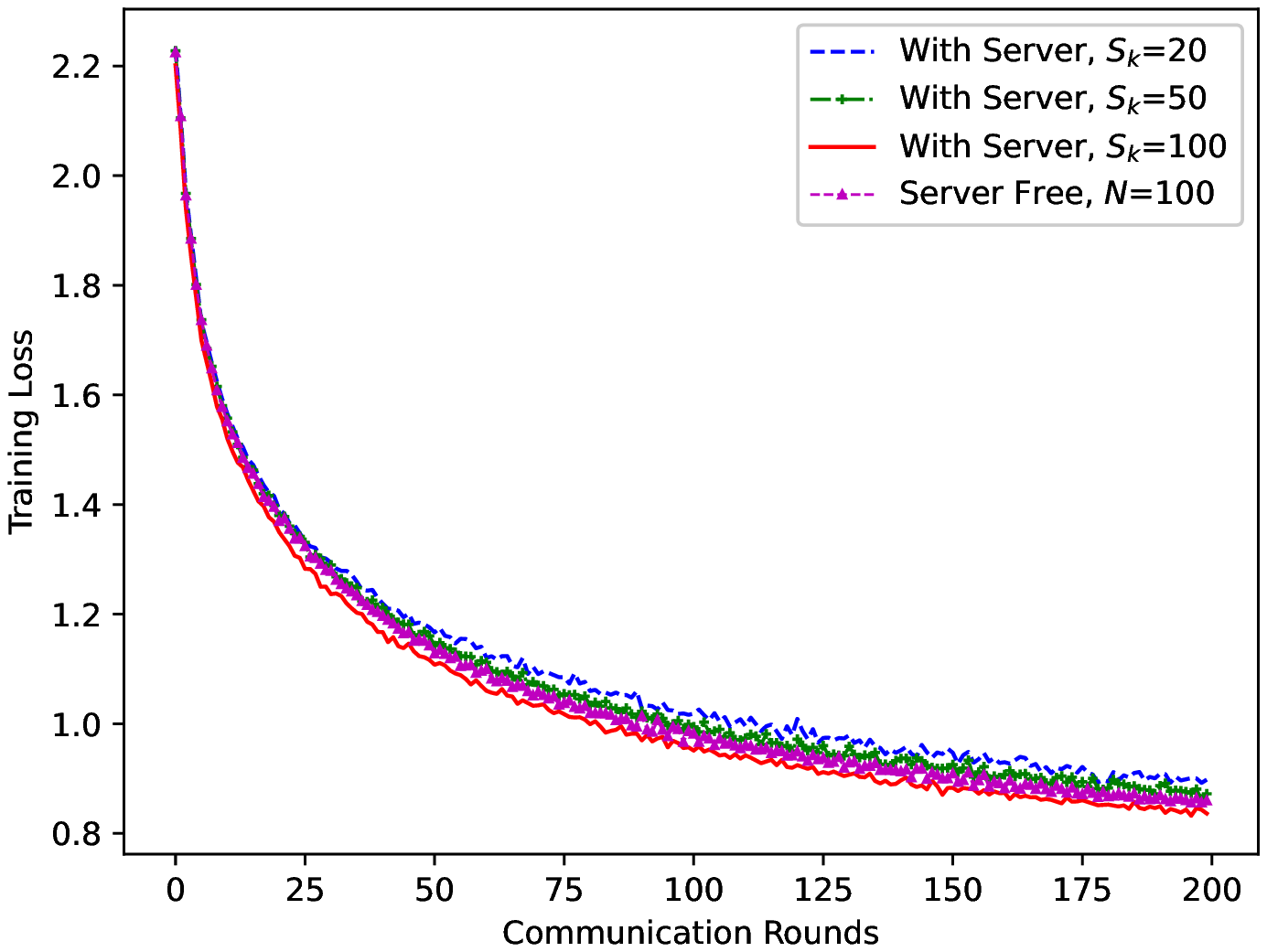}}
  \caption{ Comparison of convergence rates between SFWFL and a server-based FL: ($a$) training the MLP on the MNIST dataset and ($b$) training the CNN on the CIFAR-10 dataset.  }
  \label{fig:ComprSFWFLwFL}
\end{figure}

\begin{figure}[t!]
  \centering

  \subfigure[\label{fig:3a}]{\includegraphics[width=0.95\columnwidth]{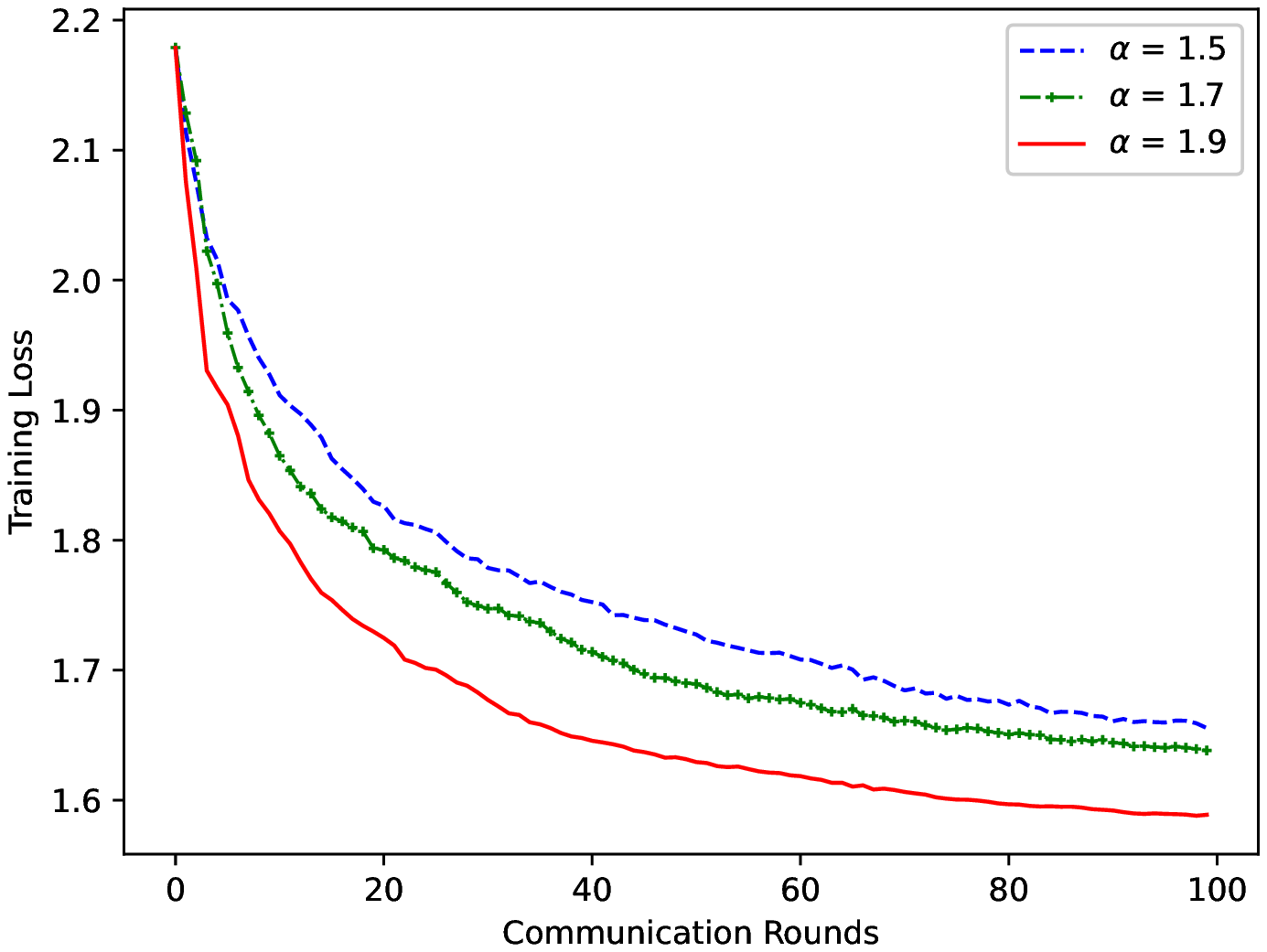}} ~
  \subfigure[\label{fig:3b}]{\includegraphics[width=0.95\columnwidth]{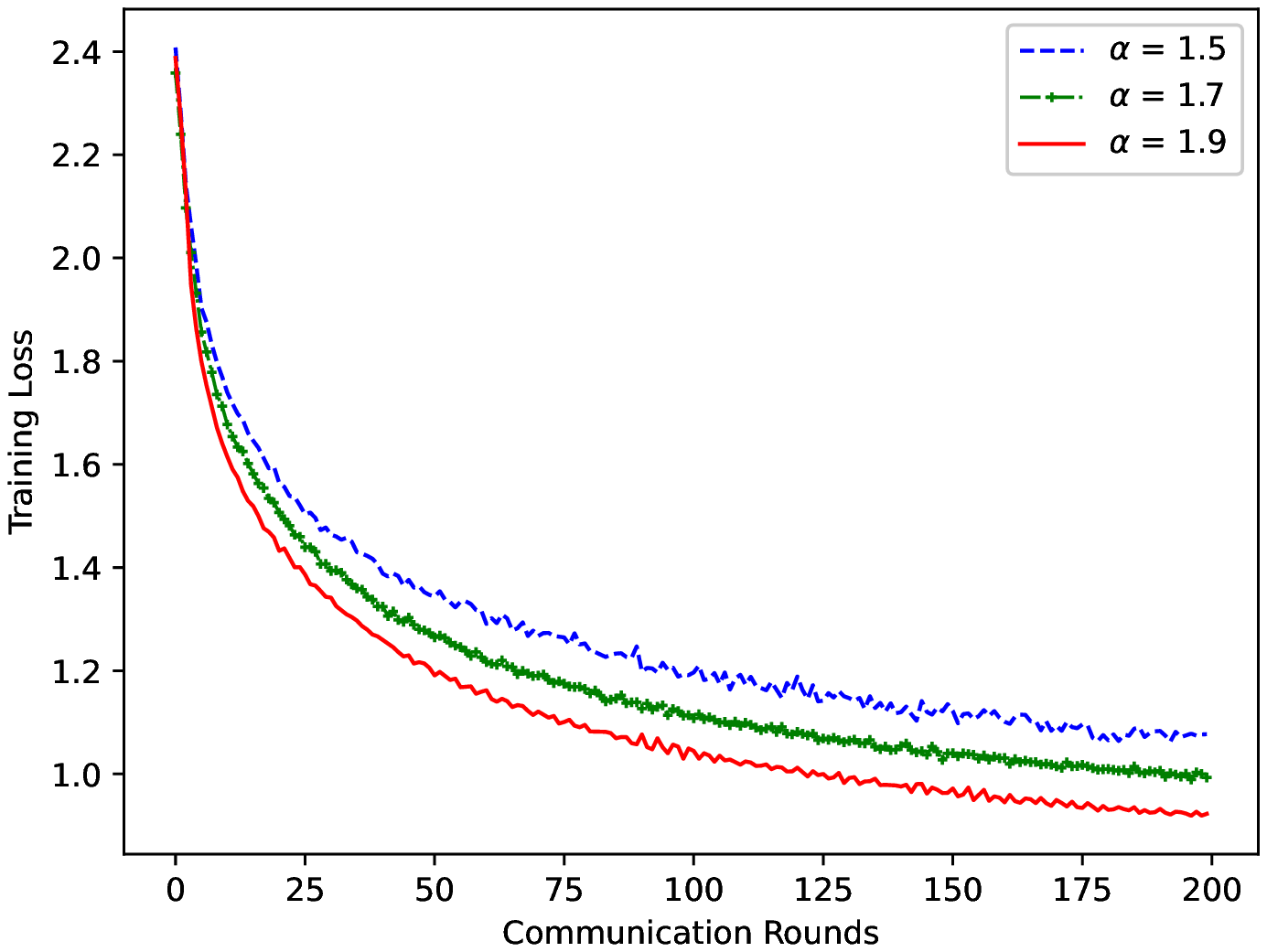}}
  \caption{ Simulation results of the training loss under SFWFL: ($a$) training the MLP on the MNIST dataset and ($b$) training the CNN on the CIFAR-10 dataset.  }
  \label{fig:AlphaEffc}
\end{figure}

Fig.~\ref{fig:ComprSFWFLwFL} compares the convergence performance between SFWFL and the conventional setting of FL based on an edge server. 
Specifically, a typical server-based FL has an architecture as Fig.~\ref{fig:FL_Systems_comparison}($a$): In each communication round $k$, an edge server collects the gradients from a subset of UEs, denoted as $\mathcal{S}_k$, to improve the global model and feeds back the update back to them. 
In this experiment, we set the total number of UEs to be $N=100$ and the tail index as $\alpha=2$, namely, the interference obeys a Gaussian distribution. 
We consider the server-based FL employs digital communication (which involves encoding/decoding and modulation/demodulation processes) to transmit the parameters. 
Hence, the UEs' gradients can be received by the server without error in the conventional FL setting. 
We vary the size of $\mathcal{S}_k$ to reflect the constraint of communication resources of the system. 
From this figure, we observe that the convergence rates of SFWFL and server-based FL with full communication resources (i.e. $\mathcal{S}_k = 100$) closely match with each other.
As such, SFWFL attains the exhaustive power of server-based FL in the absence of a costly edge server.
Another, perhaps more strikingly, message conveyed by Fig.~\ref{fig:ComprSFWFLwFL} is that, when there are insufficient communication resources, SFWFL that adopts analog transmissions (hence resulting in noisy gradients) can even outperform an FL that is based on digital communications (which promote error-free gradients) in terms of convergence rate.
Since analog transmissions can be achieved by elementary communication components such as amplitude modulation and match filtering, the conclusion from this observation seems to put in vain all the efforts we have spent in developing better communication and signal processing technologies that enhance the quality of aggregated gradients. 
The following simulation result shows that it is not necessarily true.

% \begin{figure}[t!]
%   \centering{}

%     {\includegraphics[width=0.95\columnwidth]{Figures/ServerFree_alpha.eps}}

%   \caption{ Simulation results of the training loss of MLP on the MNIST data set, under different tail index $\alpha$. }
%   \label{fig:AlphaEffc}
% \end{figure}

Fig.~\ref{fig:AlphaEffc} plots the training loss of SFWFL as a function of the communication rounds for a varying value of the tail index.
The figure shows that ($a$) the training loss decays steadily along with the communication rounds, regardless of the heaviness of the tail in the interference distribution, and ($b$) the tail index $\alpha$ plays a critical role in the rate of convergence. 
Notably, an increase in the tail index leads to a significant speedup in the convergence rate, whereas the improvement is non-linear with respect to $\alpha$. 
These observations are in line with Remark~2.  
Therefore, interference is the dominating factor in the performance of SFWFL. 
In networks with adequate interference, this algorithm has a delicate convergence rate. However, in the presence of strong interference, the convergence performance quickly deteriorates.

\begin{figure}[t!]
  \centering

  \subfigure[\label{fig:4a}]{\includegraphics[width=0.95\columnwidth]{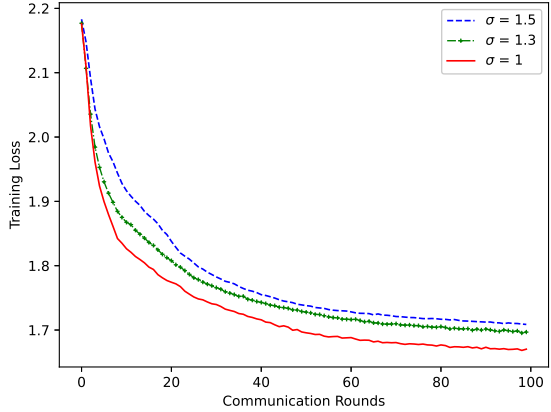}} ~
  \subfigure[\label{fig:4b}]{\includegraphics[width=0.95\columnwidth]{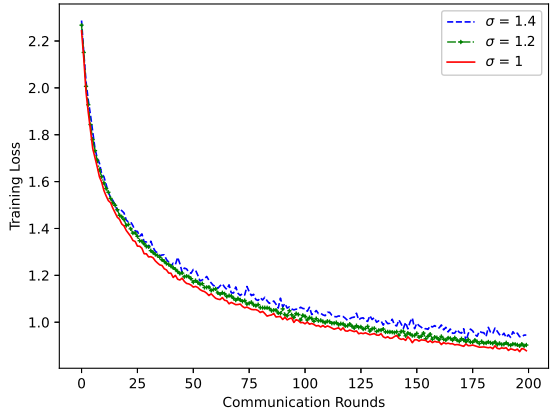}}
  \caption{ Simulation results of the training loss under SFWFL: ($a$) training the MLP on the MNIST dataset and ($b$) training the CNN on the CIFAR-10 dataset.  }
  \label{fig:FadingEffc}
\end{figure}

Because of analog transmissions, the aggregated gradients in SFWFL is perturbed by random channel fading. We investigate such an effect in Fig.~\ref{fig:FadingEffc}.
Particularly, the figure draws the convergence curve under different conditions of the channel fading. 
We observe that ($a$) a high variance in the channel fading increases the chance of encountering deep fade in the transmission, which inflicts the model training process and ($b$) unlike the effect of interference, variance in the channel fading only brings a mild influence on the convergence rate. This observation coincides with the theoretical finding in Remark~3.

Fig.~\ref{fig:PartNumEffc} evaluates the scaling effect of SFWFL, in which the training loss versus communication rounds curve is depicted under different numbers of UEs in the network. 
We can see that the algorithm's convergence rate increases with respect to $N$, confirming the conclusion in Remark~4 that enlarging the number of UEs is beneficial for the system. 
The reason can be ascribed to two crucial facts: ($a$) as each UE owns an individual dataset, an increase in $N$ allows the aggregated gradient to ingest more data information in every round of global iteration, because all the UEs can concurrently access the radio channel and upload their locally trained parameters, and ($b$) more UEs participating in the analog transmission can reduce the impact of channel fading, as explained in Remark~4. 
Nevertheless, we shall also emphasize that such an effect is less significant compared to the tail index because it only influences the multiplier in the convergence rate.

\begin{figure}[t!]
  \centering{}

    {\includegraphics[width=0.95\columnwidth]{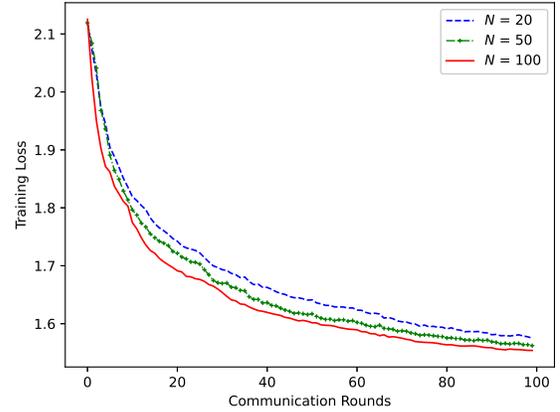}}

  \caption{ Simulation results of the training loss of MLP on the MNIST data set, under different number of UEs $N$. }
  \label{fig:PartNumEffc}
\end{figure}

Finally, we put the spotlight on the Zero-Wait SFWFL. 
Since the notion of global iteration refers to different time scales under the zero wait and compute-and-wait versions of SFWFL, we define the speedup via restless computing as follows: 
\begin{align*}
  \text{Speedup} = \frac{K(M+DM+\tau_{\mathrm{L}})}{K(M+\tau_{\mathrm{G}})}
\end{align*}
where $\tau_{\mathrm{L}}$ and $\tau_{\mathrm{G}}$ represent the local and global aggregation times, respectively. 
Then, we summarize the accuracy and run time comparison under different scales of communication latency in Table~\ref{table:Comparison_ZW_CaW}.
The results amply demonstrate that upon reaching the same accuracy, Zero-Wait SFWFL attains a substantial speedup in the high latency regime. 
Such an improvement mainly attributes to that ($a$) Zero-Wait SFWFL maintains the same communication frequency as conventional SFWFL and ($b$) the local computation is pipelined with global communication.

\begin{table*}[t!]
    
  \centering
  \begin{tabular}{lcccccccc}
  
  \toprule
  
  \multirow{2}{*}{\bf{Algorithm}} 
  & \multicolumn{2}{c}{\bf{Compute-and-Wait}} & \multicolumn{2}{c}{\bf{Zero Wait}} & \multicolumn{2}{c}{\bf{Zero Wait}} &\multicolumn{2}{c}{\bf{Zero Wait}} \\
  & \multicolumn{2}{c}{$M=5$} & \multicolumn{2}{c}{$M=5,D=1$}  & \multicolumn{2}{c}{$M=5,D=2$} & \multicolumn{2}{c}{$M=5,D=4$}  \\
  \cmidrule(){2-3} \cmidrule(){4-5}  \cmidrule(){6-7} \cmidrule(){8-9}
  
                & Accuracy   & Speedup    & Accuracy  & Speedup & Accuracy  & Speedup  & Accuracy  & Speedup       \\ \midrule

  \addlinespace[0.12cm]
  MNIST IID      & 85.9& 1$\times$ & 85.6 &1.9$\times$ &85.1& 2.9$\times$  & 85.0& 4.7$\times$  \\
  \addlinespace[0.12cm]
  MNIST non-IID  & 75.7& 1$\times$ & 75.5 &1.8$\times$ &74.7& 2.8$\times$ & 75.4& 4.6$\times$ \\
  \addlinespace[0.08cm]
  % \hline
  \addlinespace[0.04cm]
  CIFAR-10 IID   &76.1& 1$\times$ & 76.2  &2.0$\times$ &73.9&2.9$\times$ & 74.4& 4.8$\times$   \\

  \bottomrule
  \addlinespace[0.08cm]

  \end{tabular}
  \caption{Performance comparison of Zero Wait  and  Compute-and-Wait}
  \label{table:Comparison_ZW_CaW}
  \end{table*}

\subsection{Discussions}
The main takeaways from this section are outlined as follows. 

\subsubsection{Federated edge learning is readily achievable, do not procrastinate}Speaking of FL, we often identify it as one of the \textit{future} technologies.
We assume FL will only occur in wireless systems beyond 5G or near 6G, where edge computing resources may be abundantly available. 
The mere fact is that end-user devices, e.g., smartphones, tablets, or laptops, already have substantial processing powers. At the same time, most of the network edge elements, e.g., WiFi hot spots, APs, or base stations, can only perform elementary signal processing functions. 
That said, it shall not discourage us from realizing a collaboratively intelligent system from the present wireless networks.
The SFWFL developed in this paper demonstrates a new way to build large-scale FL on the currently available infrastructure. 
And surprisingly, the scheme attains a convergence performance comparable to an FL system availed with an edge server. 

\subsubsection{No free lunch is (still) the first principle}Although SFWFL removes the edge server from the overall system and can be implemented by low-cost hardware, its functionality is accomplished at the expense of ($a$) additional occupation in UEs' local memory, as they need to store both the local model parameter and accumulated gradient so as to replace the later with the globally averaged ones and remedy the former. Such an expenditure in storage becomes more critical in the Zero-Wait version of SFWFL, since the UEs need to cache multiple accumulated gradients along the local computing process; and ($b$) the potential of dreadful degradation in the performance rate, as the tail index of interference distribution is directly affecting the exponential factor of the convergence rate. The model training process may be severely slowed down in the presence of strong electromagnetic interference.  

\subsubsection{Elephant in the room}To boost communication efficiency in FL, a plethora of techniques have been developed, ranging from compressing the local parameters, pruning the computing architecture, to developing better UE scheduling policies.
The common goal of these methods is to reduce the parameter transmission time such that UEs do not need to wait for too long before they can receive the improved global model and perform a new round of local training. 
The proposed Zero-Wait SFWFL points out that the on-device computing can be executed in parallel to the parameter updating. 
Therefore, the bottleneck of wireless FL is not in the communication but the algorithm design. 
We expect disclosing this fact future can promote further research pursuits to develop better algorithms that enhance the FL system.

%%---------------------------------------------------------------------------%
%%                           Sec: Conclusion                                 %
%%---------------------------------------------------------------------------%
\section{ Conclusion }\label{sec:Conclusion}
In this paper, we established the SFWFL architecture, which exploits the superposition property of analog waveform to achieve FL training in a wireless network without using an edge server. 
Unlike the fully decentralized version of FL, we do not abandon the star connection. 
Instead, we lean on such a centralized structure for fast and scalable learning in a network with massively distributed UEs. 
The proposed SFWFL not only boasts a very high communication efficiency, but also can be implemented in low-cost hardware and has a built-in privacy enhancement.
We also improved the training algorithm to parallel the UEs' local computing with global parameter transmissions. 

In order to evaluate the developed framework, we derived the convergence rates for both SFWFL and its improved version, coined as Zero-Wait SFWFL.
The analytical results revealed that SFWFL can attain a similar, or even outperform the convergence rate of a server-based FL, if the interference has a Gaussian distribution. 
It also showed that the convergence performance of SFWFL is sensitive to the heavy tailedness of interference distribution, whereas the convergence rate deteriorates quickly as the interference tail index decreases. Yet, running local computation in concurrence with global communications is always beneficial to reducing the system run time, and the gain is particularly pronounced in high communication latency. 
These theoretical findings have been validated through excessive simulations. 

The architecture, algorithm, and analysis developed in this paper have set up a new distributed learning paradigm 
in which many future researches can nest and grow.
For instance, one can explore the effects of adopting multiple antennas at the AP, which can be used to manoeuvre power boosting and/or interference cancellation \cite{YanGerQue:17}, on the performance of SFWFL.  
Investigating the impacts of UEs' mobility on the system's performance is also a concrete direction \cite{FenYanHu:21TWC}. 
Another future extension of the present study is to reduce the sensitivity of SFWFL to heavy-tailed interference via, e.g., the gradient clipping schemes \cite{GorDanGas:20}

%%---------------------------------------------------------------------------%
%%                           Sec: Appendix                                   %
%%---------------------------------------------------------------------------%
\begin{appendix}
% ============================= %
%      Theorem 1: Proof 
% ============================= %
\subsection{Proof of Theorem~\ref{thm:ConvAnals}} \label{Apndx:ConvAnals_proof}
For ease of exposition, we denote $\Delta_k = \boldsymbol{w}_k - \boldsymbol{w}^*$. 
According to \eqref{equ:ZW_Intliz}, we have 
\begin{align}
\boldsymbol{w}_{k+1} = \boldsymbol{w}_k - \eta_k \Big( \frac{1}{N} \sum_{ n = 1 }^N h_{n,k} \bar{ \boldsymbol{g} }^n_k + \boldsymbol{\xi}_k \Big).
\end{align}
Then, using Lemma~1, we can expand the expectation on the $\alpha$-moment of $\Delta_{k+1}$ via the following:
\begin{align} \label{equ:Delta_Bound}
&\mathbb{E} \Big[ \big\Vert \Delta_{k+1} \big\Vert_\alpha^\alpha \Big] = \mathbb{E} \Big[ \Big\Vert \Delta_{k} - \eta_k \big( \frac{1}{N} \sum_{ n = 1 }^N h_{n,k} \bar{ \boldsymbol{g} }^n_k + \boldsymbol{\xi}_k \big) \Big\Vert_\alpha^\alpha \Big]
\nonumber\\
&\leq \underbrace{ \mathbb{E} \Big[ \big\Vert \Delta_k -  \eta_k M \nabla f( \boldsymbol{w}_k ) \big\Vert_\alpha^\alpha \Big] }_{Q_1}  
\nonumber\\
&+ 4 \eta_k^\alpha \underbrace{ \mathbb{E} \Big[ \big\Vert M \nabla f( \boldsymbol{w}_k ) - \frac{1}{N} \sum_{ n = 1 }^N \bar{ \boldsymbol{g} }^n_k \big\Vert_\alpha^\alpha \Big] }_{Q_2}  
\nonumber\\
&+ 4 \eta_k^\alpha \underbrace{ \mathbb{E} \Big[ \big\Vert \frac{1}{N} \sum_{ n = 1 }^N \big( h_{n,k} - 1 \big) \bar{ \boldsymbol{g} }^n_k \big\Vert_\alpha^\alpha \Big] }_{Q_3}  +\, 4 C \eta_k^\alpha. 
\end{align}

By leveraging Lemma~2, we bound $Q_1$ as follows:
\begin{align} \label{equ:Q1_bound}
Q_1 &\stackrel{(a)}{=}  \mathbb{E} \Big[ \big\Vert \Delta_k - \eta_k M \big(\, \nabla f ( \boldsymbol{w}_k ) - \nabla f ( \boldsymbol{w}^* ) \,\big) \big\Vert_\alpha^\alpha \Big]
\nonumber\\
&= \mathbb{E} \Big[ \big\Vert \Delta_k - \mu \eta_k \nabla^2 f ( \boldsymbol{w}^{\sharp}_k ) \Delta_k \big\Vert_\alpha^\alpha \Big]
\nonumber\\
& \leq \Big\Vert \boldsymbol{I}_d - \eta_k M \nabla^2 f ( \boldsymbol{w}^{\sharp}_k ) \Big\Vert_\alpha^\alpha \times \mathbb{E} \Big[ \big\Vert \Delta_k \big\Vert_\alpha^\alpha \Big]
\nonumber\\
& \stackrel{(a)}{=}  ( 1 - \eta_k M L ) \times \mathbb{E} \Big[ \big\Vert \Delta_k \big\Vert_\alpha^\alpha \Big]
\end{align}
where ($a$) holds because $\nabla f ( \boldsymbol{w}^* ) = 0$.

Next, we establish the following bound for $Q_2$: 
\begin{align} \label{equ:Q2_bound}
& Q_2 = \mathbb{E} \Big[ \big\Vert \frac{1}{N} \sum_{n=1}^N \sum_{ i=1 }^M \big( \nabla f_n ( \boldsymbol{w}^n_{k,i}; \gamma_i ) - \nabla f_n ( \boldsymbol{w}_k ) \big) \big\Vert_\alpha^\alpha \Big]
\nonumber\\
&\stackrel{(a)}{ \leq } \! d^{ 1 - \frac{1}{\alpha} } \! \cdot \! \mathbb{E} \Big[ \Big( \big\Vert \frac{1}{N} \! \sum_{n=1}^N \sum_{ i=1 }^M \! \big( \nabla f_n ( \boldsymbol{w}^n_{k,i}; \gamma_i ) -\! \nabla f_n ( \boldsymbol{w}_k ) \big) \big\Vert_2^2 \Big)^{ \!\! \frac{ \alpha }{ 2 } }  \Big]
\nonumber\\
&\stackrel{(b)}{ \leq } \! d^{ 1 - \frac{1}{\alpha} } \! \cdot \!  \Big( \frac{1}{N} \! \sum_{n=1}^N \mathbb{E} \Big[ \big\Vert  \sum_{ i=1 }^M \! \big( \nabla f_n ( \boldsymbol{w}^n_{k,i}; \gamma_i ) -\! \nabla f_n ( \boldsymbol{w}_k ) \big) \big\Vert_2^2 \Big] \Big)^{ \! \frac{ \alpha }{ 2 } }  
\nonumber\\
&\stackrel{(c)}{ \leq } \! d^{ 1 - \frac{1}{\alpha} } \! \cdot \!  \Big( \frac{ \lambda^2 M }{N} \! \sum_{n=1}^N \sum_{ i=1 }^M \mathbb{E} \Big[ \big\Vert   \boldsymbol{w}^n_{k,i} - \boldsymbol{w}_k  \big\Vert_2^2 \Big] \Big)^{ \! \frac{ \alpha }{ 2 } }  
\nonumber\\
&= d^{ 1 - \frac{1}{\alpha} } \! \cdot \!  \Big( \frac{ \lambda^2 M }{N} \! \sum_{n=1}^N \sum_{ i=1 }^M \mathbb{E} \Big[ \big\Vert \eta_k \sum_{ j = 1 }^i \nabla f_{n} ( \boldsymbol{w}^n_{k,j}; \gamma_j )  \big\Vert_2^2 \Big] \Big)^{ \! \frac{ \alpha }{ 2 } }  
\nonumber\\
& \leq d^{ 1 - \frac{1}{\alpha} } \lambda^\alpha M^{ 2 \alpha } G^\alpha
\end{align}
in which ($a$) and ($b$) follow from the Holder's inequality and Jensen's inequality, respectively, and ($c$) is owing to smoothness property of $f_n(\cdot)$ as per Assumption~2.

In a similar vein, $Q_3$ can be bounded as:
\begin{align} \label{equ:Q3_bound}
& Q_3 = \frac{ d^{ 1 - \frac{1}{\alpha} } }{ N^{ \alpha } } \cdot  \mathbb{E} \Big[ \Big( \big\Vert \sum_{n=1}^N  \big( h_{ n,k } - 1 \big) \bar{ \boldsymbol{g} }^n_k \big\Vert_2^2 \Big)^{ \!\! \frac{ \alpha }{ 2 } }  \Big]
\nonumber\\
& \leq \frac{ \sigma^\alpha d^{ 1 - \frac{1}{\alpha} } }{ N^{ \alpha } } \cdot \Big( \sum_{n=1}^N \mathbb{E} \Big[ \big\Vert \sum_{ i = 1 }^M \nabla f_{n} ( \boldsymbol{w}^n_{k,i}; \gamma_i )  \big\Vert_2^2 \Big] \Big)^{ \! \frac{ \alpha }{ 2 } }  
\nonumber\\
& \leq \frac{ \sigma^\alpha M^\alpha G^\alpha d^{ 1 - \frac{1}{\alpha} } }{ N^{ \alpha / 2 } }. 
\end{align}

Finally, we substitute \eqref{equ:Q1_bound}, \eqref{equ:Q2_bound}, and \eqref{equ:Q3_bound} into \eqref{equ:Delta_Bound}, which results in:
\begin{align} \label{equ:DeltatFnalBnd}
&\mathbb{E} \Big[ \big\Vert \Delta_{k+1} \big\Vert_\alpha^\alpha \Big] \leq \left( 1 - \eta_k M L \right) \times \mathbb{E} \Big[ \big\Vert \Delta_k \big\Vert_\alpha^\alpha \Big] 
\nonumber\\
&\qquad \qquad  + 4 \eta_k^\alpha \Big( C + d^{ 1 - \frac{1}{\alpha} } \lambda^\alpha G^\alpha M^{ 2 \alpha } + \frac{ \sigma^\alpha G^\alpha M^\alpha d^{ 1 - \frac{1}{\alpha} } }{ N^{\alpha/2} } \Big). 
\end{align}
The proof is completed by invoking Lemma~3 in \cite{YanCheQue:21JSTSP} to the above inequality.

% ============================= %
%        Lemma 3: Proof 
% ============================= %
\subsection{Proof of Lemma~\ref{lma:PrmtUnivBnd} } \label{Apndx:PrmtUnivBnd_proof}
Following the update process of local parameters in Algorithm~2, we have 
\begin{align}
\boldsymbol{w}^n_k = \bar{ \boldsymbol{w} }_{ k-D } &- \eta_{ k - D } \, \bar{ \boldsymbol{g} }^n_{ k - D } 
\nonumber\\
&- \eta_{ k - D + 1 } \, \bar{ \boldsymbol{g} }^n_{ k - D + 1 } - \cdots - \eta_{ k - 1 } \, \bar{ \boldsymbol{g} }^n_{ k - 1 }.
\end{align}
For any two UEs $m$ and $n$, the following holds:
\begin{align}
&\mathbb{E} \big[ \Vert \boldsymbol{w}^n_k - \boldsymbol{w}^m_k \Vert^2_2 \big] 
\nonumber\\
&= \mathbb{E} \big[ \Vert \eta_{ k - D } ( \bar{ \boldsymbol{g} }^n_{ k - D } - \bar{ \boldsymbol{g} }^m_{ k - D }) + \cdots +  \eta_{ k - 1 } ( \bar{ \boldsymbol{g} }^n_{ k - 1 } - \bar{ \boldsymbol{g} }^m_{ k - 1 }) \Vert^2_2 \big]
\nonumber\\
&\stackrel{(a)}{ \leq } 4 \eta^2_{ k - D } M^2 D^2 G^2 
\end{align}
where ($a$) follows from the fact that the local parameters of UE $m$ and UE $n$ differ by at most $DM$ SGD iterations. 
The proof then follows by using Jensen's inequality in the following way:
\begin{align}
 & \mathbb{E} \big[ \Vert \boldsymbol{w}^n_k - \bar{\boldsymbol{w}}_k \Vert^2_2 \big] = \mathbb{E} \big[ \Vert \boldsymbol{w}^n_k - \frac{1}{N} \sum_{ m=1 }^N \boldsymbol{w}^m_k \Vert^2_2 \big] 
\nonumber\\
&= \mathbb{E} \big[ \Vert  \frac{1}{N} \sum_{ m=1 }^N \big( \boldsymbol{w}^n_k - \boldsymbol{w}^m_k \big) \Vert^2_2 \big] 
\nonumber\\
& \leq  \frac{1}{N} \sum_{ m=1 }^N \mathbb{E} \big[ \Vert  \big( \boldsymbol{w}^n_k - \boldsymbol{w}^m_k \big) \Vert^2_2 \big] \leq 4 \eta^2_{ k - D } M^2 D^2 G^2.
\end{align}

% ============================= %
%        Theorem 2: Proof 
% ============================= %
\subsection{Proof of Theorem~\ref{thm:ZW_SFWFL_ConvAnals}} \label{Apndx:ZW_SFWFL_ConvAnals_proof}
Similar to the proof of Theorem~1, we denote by $\bar{\Delta}_k = \bar{ \boldsymbol{w}}_k - \boldsymbol{w}^*$. 
The model training procedure in Algorithm~2 stipulates the following relationship:
\begin{align}
\bar{\Delta}_{ k + 1 } = \bar{\Delta}_k - \eta_k \sum_{ n=1 }^N \bar{ \boldsymbol{g} }^n_k + \eta_{ k - D } \Big( \frac{ 1 }{ N } \sum_{ n = 1 }^N \bar{ \boldsymbol{g} }^n_{ k - D } - \tilde{ \boldsymbol{g} }_{ k - D } \Big).
\end{align}
Taking similar steps as \eqref{equ:Delta_Bound}, we arrive at the following:
\begin{align} \label{equ:Delta_Bar_Bound}
&\mathbb{E} \Big[ \big\Vert \bar{ \Delta }_{k+1} \big\Vert_\alpha^\alpha \Big] 
\leq \underbrace{ \mathbb{E} \Big[ \big\Vert \bar{ \Delta }_k -  \eta_k M \nabla f( \bar{ \boldsymbol{w} }_k ) \big\Vert_\alpha^\alpha \Big] }_{Q_4}  
\nonumber\\
&+ 4 \eta_{ k - D }^\alpha \underbrace{ \mathbb{E} \Big[ \big\Vert \frac{1}{N} \sum_{ n = 1 }^N \big( h_{n,k - D} - 1 \big) \bar{ \boldsymbol{g} }^n_{ k - D } \big\Vert_\alpha^\alpha \Big] }_{Q_5} 
\nonumber\\
&+ 4 \eta_k^\alpha \underbrace{ \mathbb{E} \Big[ \big\Vert  \frac{1}{N} \sum_{ n = 1 }^N \big(\, \bar{ \boldsymbol{g} }^n_k - M \nabla f( \boldsymbol{w}_k ) \big)  \big\Vert_\alpha^\alpha \Big] }_{Q_6}    +\, 4 C \eta_{ k - D }^\alpha. 
\end{align}
Using the results in Theorem~1, we have
\begin{align}
Q_4 \leq ( 1 - \eta_k M L ) \times \mathbb{E} \Big[ \big\Vert \bar{\Delta}_k \big\Vert_\alpha^\alpha \Big]
\end{align}
and 
\begin{align}
Q_5 \leq \frac{ \sigma^\alpha M^\alpha G^\alpha d^{ 1 - \frac{1}{\alpha} } }{ N^{ \alpha / 2 } }.
\end{align}
The quality $Q_6$ can be bounded as follows:
\begin{align}
& Q_6 = \mathbb{E} \Big[ \big\Vert \frac{1}{N} \sum_{n=1}^N \sum_{ i=1 }^M \big( \nabla f_n ( \boldsymbol{w}^n_{k,i}; \gamma_i ) - \nabla f_n ( \bar{ \boldsymbol{w} }_k ) \big) \big\Vert_\alpha^\alpha \Big]
\nonumber\\
& \leq  d^{ 1 - \frac{1}{\alpha} } \! \cdot \!  \Big( \frac{1}{N} \! \sum_{n=1}^N \mathbb{E} \Big[ \big\Vert  \sum_{ i=1 }^M \! \big( \nabla f_n ( \boldsymbol{w}^n_{k,i}; \gamma_i ) -\! \nabla f_n ( \bar{ \boldsymbol{w} }_k ) \big) \big\Vert_2^2 \Big] \Big)^{ \! \frac{ \alpha }{ 2 } }  
\nonumber\\
&\stackrel{(a)}{ \leq } \! d^{ 1 - \frac{1}{\alpha} } \! \cdot \!  \Big( \frac{ \lambda^2 M }{N} \! \sum_{n=1}^N \sum_{ i=1 }^M \mathbb{E} \Big[ \big\Vert   \boldsymbol{w}^n_{k,i} - \boldsymbol{w}_k  \big\Vert_2^2 \Big] \Big)^{ \! \frac{ \alpha }{ 2 } }  
\nonumber\\
& \leq d^{ 1 - \frac{1}{\alpha} } \lambda^\alpha 2^\alpha ( 1 + D )^\alpha G^\alpha M^{ 2 \alpha }
\end{align}
where ($a$) results from Lemma~3. Putting the above bounds into \eqref{equ:Delta_Bar_Bound}, we obtain the following:
\begin{align} \label{equ:BarDeltatFnalBnd}
&\mathbb{E} \Big[ \big\Vert \bar{ \Delta }_{k+1} \big\Vert_\alpha^\alpha \Big] \leq \left( 1 - \eta_k M L \right) \times \mathbb{E} \Big[ \big\Vert \bar{ \Delta }_k \big\Vert_\alpha^\alpha \Big] + 4 \eta_{k-D}^\alpha C
\nonumber\\
& + 4 \eta_{ k }^\alpha d^{ 1 - \frac{1}{\alpha} } \lambda^\alpha 2^\alpha ( 1 \!+\! D )^\alpha G^\alpha \! M^{ 2 \alpha }
% \nonumber\\
% &\qquad \qquad  
 \!+\! 4 \eta_{k-D}^\alpha \frac{ \sigma^\alpha G^\alpha \! M^\alpha d^{ 1 - \frac{1}{\alpha} } }{ N^{\alpha/2} }. 
\end{align}
By assigning $\eta_k = \theta/k$, we have $\eta_{ k - D } \approx \eta_k$ for $k \gg 1$.
Then, Theorem~2 follows by applying Lemma~3 of \cite{YanCheQue:21JSTSP} to the above.

\end{appendix}
%\balance
\bibliographystyle{IEEEtran}
\bibliography{bib/StringDefinitions,bib/IEEEabrv,bib/howard_SFWFL}

\end{document}